\documentclass[11pt]{article}

\usepackage{booktabs}
\usepackage{array}
\usepackage{longtable}
\usepackage{multirow}
\usepackage{graphicx}
\graphicspath{{./}{../}}
\usepackage{bm}
\usepackage{xcolor}
\usepackage{tikz}
\usetikzlibrary{arrows.meta,positioning,fit,decorations.pathreplacing}
\usepackage{amsmath,amssymb}
\usepackage{float}
\usepackage{titlesec}
\usepackage[normalem]{ulem}
\usepackage[nodisplayskipstretch]{setspace}
\usepackage{etoolbox}
\usepackage[hang, flushmargin]{footmisc}
\usepackage[round]{natbib}
\bibpunct{(}{)}{;}{a}{,}{,}
\usepackage[colorlinks=true, linkcolor=blue, citecolor=black, urlcolor=blue]{hyperref}

\newcommand{\sym}[1]{\ifmmode^{\text{#1}}\else\textsuperscript{#1}\fi}

\titleformat*{\section}{\Large\bfseries}
\titleformat*{\subsection}{\large\bfseries}
\titleformat*{\paragraph}{\large\bfseries}
\titleformat*{\subparagraph}{\large\bfseries}

\usepackage[margin=1.0in]{geometry}

\usepackage[parfill]{parskip}
\renewcommand{\footnoterule}{
  \kern 0pt
  \hrule width 1.0\textwidth height 0.4pt
  \kern 2.5pt
}

\AtBeginEnvironment{table}{\singlespacing}
\AtBeginEnvironment{figure}{\singlespacing}
\AtBeginEnvironment{longtable}{\singlespacing}

\begin{document}

\title{\LARGE Human Capital, AI, and Labor Commoditization}
\author{
Auyon Siddiq \qquad Niuniu Zhang\\[0.5em]
\textit{UCLA Anderson School of Management}
}
\date{June 19, 2026}

\maketitle
\begingroup
\renewcommand\thefootnote{}
\footnotetext{Contact: Auyon Siddiq, \texttt{auyon.siddiq@anderson.ucla.edu}; Niuniu Zhang, \texttt{niuniu.zhang.phd@anderson.ucla.edu}.}
\endgroup

\begin{spacing}{1.05}
\begin{abstract}
\noindent Has generative AI changed how labor markets value human capital? We study this question using contract-level data from Upwork, a large online labor market. We represent worker profiles with high-dimensional text embeddings, allowing us to capture rich human capital information from unstructured profile text. We then compute the predictive importance of workers' human capital information and posted hourly rates for client demand, and incorporate these measures into a difference-in-differences design around the release of ChatGPT. We find that in more AI-exposed job categories, the importance of human capital declines and the importance of price rises, suggesting a commoditization effect of AI on labor. Two additional findings support commoditization as a mechanism: The demand premium enjoyed by workers with strong human capital declines in more AI-exposed categories, and demand reallocates toward lower-priced workers. Our results have implications for the design of online labor markets, workers' incentives to invest in human capital, and labor welfare. \\

\end{abstract}
\end{spacing}

\newpage

\section{Introduction}

Human capital, including education, experience, and technical skills, shapes workers' productivity and earnings \citep{becker1964,mincer1974}. Information about workers' human capital also helps labor markets function by allowing employers to form beliefs about worker productivity prior to hiring \citep{spence1973,stiglitz1975,altonji2001}. This reliance is especially pronounced in online labor markets, where clients contract for discrete jobs from a distance and often without prior interactions with the worker. To facilitate matching and hiring, online labor markets typically organize workers' human capital information along several dimensions, including titles, descriptions, formal education, employment history, skills,  on-platform ratings, and prior client feedback, which reduce information asymmetry and help clients differentiate among workers \citep{pallais2014,stanton2016,kokkodis2016}.

The human capital attributes, valued by labor markets, may evolve as workers adopt new technologies, which alter workers' tasks and output \citep{autor2003,acemoglu2011,acemoglu2019automation}. In this regard, generative AI represents a potentially unprecedented shock to the role of human capital in labor markets. Empirical evidence suggests that AI use can compress output differences across workers by raising the performance of lower-skilled workers more than that of higher-skilled workers \citep{noy2023,peng2023,dellacqua2023,brynjolfsson2025}. This compression of worker productivity has been theorized to {\it commoditize labor}, leading employers to view workers with different levels of human capital as relatively more substitutable \citep{fukui2026}. Although labor commoditization has been proposed theoretically \citep{fukui2026}, to our knowledge, prior work has not empirically tested for its presence in a real labor market. To that end, this paper asks: {\it Has generative AI changed the role of price and human capital in the allocation of demand? Is there evidence of AI-based labor commoditization?}

We study this question using detailed data from Upwork, a large online labor market. We construct a panel of 49,610 workers who completed 2.26 million contracts from January 2021 to March 2026, a period spanning the release of ChatGPT in November 2022 \citep{openai2022chatgpt}. Our data focuses on worker attributes and employment outcomes -- including multiple signals of human capital from workers' profiles, each worker's posted hourly rate, and contract-level observations of hirings. We do not observe data related to job postings or workers' proposals. In this regard, our data is closer to prior work that has focused on hiring outcomes \citep{hui2024}, rather than the hiring process \citep{demirci2025, cowgill2026does}.

A central empirical challenge is that much of what constitutes human capital is expressed in unstructured text in the worker's profile. A standard empirical approach would involve extracting a set of variables from this information, such as years of education or indicators for particular skills. However, doing so requires researchers to decide ex ante which human capital features are likely to matter and how to encode them, risking the loss of valuable information \citep{vafa2025estimating}. We take a different approach: Rather than hand-code the content of worker profiles, we represent profile text using dense vector embeddings from a pre-trained language model. This preserves the high-dimensional content of workers' profiles without requiring us to pre-specify the most relevant concepts or words. Our method is also motivated by prior work showing that unstructured text can inform demand estimation, platform search, and matching in online markets \citep{archak2011deriving,ghose2019modeling,xu2021physician,hong2021direct,marinescu2020opening,wiles2025}.

Our approach involves first predicting individual worker demand from human capital signals and price to approximate the market allocation of demand across workers. At the worker level, we use Shapley regression values \citep{lundberg2017} to measure the importance of different human capital signals and price for predicted demand. We then incorporate these importance measures in a difference-in-differences design around the release of ChatGPT, using the occupational AI exposure scores of \citet{eloundou2024} as a continuous treatment, to test whether generative AI shifted the weight the market places on different signals of workers' human capital and price.

Our main findings are as follows. After the release of ChatGPT, the importance of human capital signals for predicted demand declines more for workers in more AI-exposed categories. Relative to a fully unexposed job category, the combined importance of all human capital signals in the most exposed job categories falls by approximately 7.8\%. Price moves in the opposite direction, rising by approximately 1.1\%. These effects grow with time and are largest toward the end of the study period, suggesting the market has not yet re-equilibrated. In short, demand for workers in AI-exposed job categories became less tied to human capital signals and more tied to the price they charge. 

Although suggestive, this observed re-weighting away from human capital signals toward price does not alone establish labor commoditization; other mechanisms such as stronger quality-signaling may also explain the increased salience of price. We therefore conduct two additional analyses to test for the plausibility of commoditization, both of which offer supporting evidence. First, we find that the demand premium enjoyed by workers with strong human capital signals declines more in high AI-exposure job categories. Second, demand shifts more aggressively from higher-price to lower-price workers in AI-exposed categories. Together, these results are consistent with AI-based labor commoditization, and suggest that clients in the market may internalize that AI standardizes worker output, leading them to down-weight human capital in favor of price when hiring. These findings have implications for the design of online labor markets, workers' incentives to invest in human capital, and labor welfare.

\subsection{Related Literature}

\textit{Generative AI and labor markets.} Our paper contributes to a growing literature that studies the effects of generative AI on labor markets. Early work identifies occupations and tasks most exposed to AI \citep{felten2021,acemoglu2022ai,eloundou2024,acemoglu2025}. Related work shows that AI can raise productivity while compressing performance differences across workers with varying stocks of human capital \citep{noy2023,peng2023,dellacqua2023,brynjolfsson2025}. Generative AI may also change task allocation and the nature of knowledge work \citep{hoffmann2024generative}. AI assistance may also reshape skill formation over time \citep{bondi2026skill,siderius2026use,aouad2026human}. Our work is closest to \citet{hui2024} and \citet{demirci2025}, both of which also study online labor markets. \citet{hui2024}  find that workers in AI-exposed occupations experience reductions in employment and earnings after the release of AI tools, with larger effects for top workers, and \citet{demirci2025} find that ChatGPT reduced demand-side job postings for automation-prone jobs.

This literature focuses primarily on aggregate market outcomes, such as labor demand and welfare. We extend it by asking whether generative AI changes the worker {\it attributes} valued by the market; specifically, we estimate how the association between labor demand and both human capital signals and price changes following the release of generative AI tools. This question is closely related to the theoretical model of labor commoditization proposed in \citet{fukui2026}, in which technical change weakens the association between human capital and productivity by standardizing worker output, making labor more commodity-like in the eyes of employers. We empirically test for this mechanism by measuring whether the predictive importance of human capital and price changes in more AI-exposed categories.

\textit{Signals of human capital in online labor markets.} A central idea in labor economics is that human capital helps explain earnings and employment \citep{becker1964,mincer1974}. \citet{lazear2009} formalizes human capital as a bundle of skills whose value depends on the weights the market attaches to each skill. Technological change can alter the organization of work and the skills that labor markets reward \citep{autor2003,acemoglu2011,acemoglu2019automation,dixon2021robot}. Because employers cannot observe worker productivity prior to hiring, employers may rely on observable signals of human capital, such as education, experience, prior performance, and reputation, to infer quality \citep{spence1973,stiglitz1975,altonji2001}.

Online labor markets make these signals both visible and consequential \citep{horton2010,horton2017}. Workers' reputation and work experience affect employment outcomes \citep{pallais2014,lin2018}. Intermediaries can further reduce information asymmetry by certifying inexperienced workers \citep{stanton2016}, and the value of reputation depends on the category and skill context in which it was earned \citep{kokkodis2016}. These platform-generated signals also shape worker behavior by mediating access to future opportunities \citep{xu2023interplay,rahman2021invisible}. \citet{kokkodis2023learning} show that clients learn to trade off reputation and price, initially experimenting with cheaper, lower-reputation workers and then shifting toward more reputable workers after unsuccessful outcomes. Other work shows that text information can affect job matching and wages \citep{marinescu2020opening}, and that skill positioning affects online labor-market outcomes \citep{fu2022,kokkodis2023}. We build on this literature by studying whether generative AI changes how human capital signals shape the market's allocation of demand across workers.

\textit{Generative AI and market information.} Related work studies how generative AI changes communication in labor markets. \citet{wiles2025} show that algorithmic writing assistance on resumes increases hiring. In contrast, \citet{cowgill2026does} show that AI can reduce the accuracy of screening workers by improving the apparent quality of their cover letters and compressing differences between high- and low-expertise senders. \citet{cui2025signaling} study an AI cover-letter tool in an online labor market and find that AI increases cover-letter tailoring but weakens the relationship between tailoring and callbacks; similarly \citet{galdin2025making} show that AI reduces the premium of customizing cover letters, as customization becomes cheaper. On the demand side, \citet{wiles2025matching} show that AI-written job-post drafts make job posts more generic and less informative to jobseekers. \cite{xu2026aiselfpreferencingalgorithmichiring} shows that LLM-based hiring screens favor resumes generated by the same model, creating a new source of bias in evaluating worker self-presentation.

This literature has focused primarily on how generative AI changes the information content of human capital signals that are primarily worker-authored. Our paper extends the analysis to more persistent signals of human capital, including education, employment history, and on-platform reputation. We also study price alongside profile information, which allows us to test whether a decline in the importance of human capital signals coincides with increased employer price sensitivity.

\section{Setting and Data}
\label{sec:setting}

\subsection{The Online Labor Market}

Our empirical setting is Upwork, a large online labor market that matches clients with workers for short-term contracts in categories spanning writing, software development, design, accounting, customer service, and administrative support. Hiring can begin from either side of the market: Clients post jobs and receive proposals from interested workers, or they search worker profiles directly and invite specific workers to apply. Contracts are either hourly or fixed price; in our sample, 61\% of contracts are fixed price and 39\% are hourly. Before hiring, clients observe several profile components, including the worker's title, written description, and skill tags used to describe the worker's services; education, employment history, and portfolio items that describe prior experience; on-platform ratings and work history; and the worker's posted hourly rate. When a contract ends, the client may leave a score and written feedback, which become part of the worker's public history, and the platform awards badges (e.g., ``Top Rated'') based on workers' performance records. 

An online labor market is particularly well suited to our research question for three reasons. First, work on the platform is organized around many short, task-based contracts, providing a direct measure of market demand for each worker. Second, in addition to the worker's posted hourly rate, the platform records rich information about the human capital signals available to clients before hiring, which we also observe. Third, clients choose among many plausibly substitutable workers who differ in their human capital signals and posted prices. Together, these features suggest that changes in what the market values should be both measurable and reflected in the allocation of demand across workers relatively quickly after the AI shock, allowing us to test for a potential commoditizing effect on labor. 

\subsection{Worker Sample}

\begin{figure}[H]
\centering
\caption{Sample Construction.}
\label{fig:sample-eligibility}
\vspace{0.75em}
\begin{tikzpicture}[
  include/.style={draw=black!70, fill=gray!8, align=center, font=\scriptsize, text width=4.75cm, minimum height=0.82cm, inner sep=4pt},
  exclude/.style={draw=black!50, fill=gray!3, align=left, font=\scriptsize, text width=4.25cm, minimum height=0.72cm, inner sep=4pt},
  arrow/.style={-{Stealth[length=2mm]}, thick, draw=black!65}
]
\node[include] (snapshot) at (0,0) {Searchable worker profiles\\(March 2026) \\\textbf{201,857 workers}};
\node[include] (active) at (0,-1.45) {$\ge 1$ contract opened in 2022\\\textbf{50,598 workers}};
\node[include] (matched) at (0,-2.90) {$\ge 1$ job subcategory recorded\\\textbf{49,610 workers}};
\node[include] (panel) at (0,-4.35) {Balanced worker-quarter panel\\2021Q1--2026Q1\\\textbf{1,041,810 worker-quarters}};

\node[exclude] (inactive) at (5.7,-0.75) {Excluded: No 2022 contracts\\\textbf{151,259 workers}};
\node[exclude] (unmatched) at (5.7,-2.20) {Excluded: No subcategory recorded\\\textbf{988 workers}};

\draw[arrow] (snapshot.south) -- (active.north);
\draw[arrow] (active.south) -- (matched.north);
\draw[arrow] (matched.south) -- (panel.north);
\draw[arrow] (snapshot.east) -- (inactive.west);
\draw[arrow] (active.east) -- (unmatched.west);
\end{tikzpicture}
\end{figure}
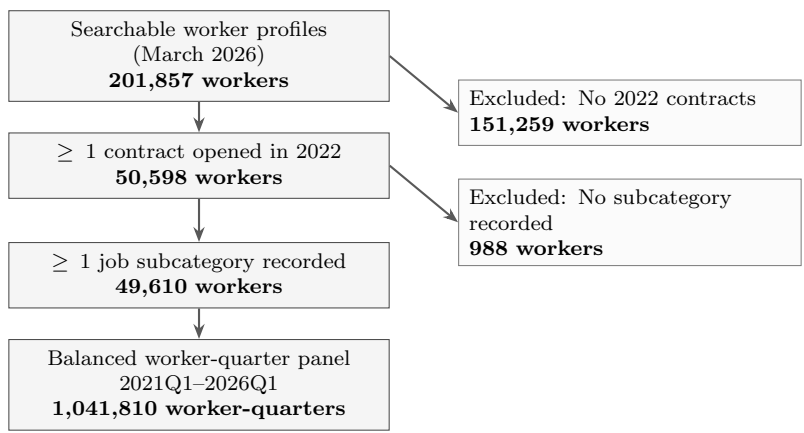

The sample is drawn from a March 2026 snapshot of worker profiles. We collect information on all searchable worker profiles in the market, resulting in 201,857 profiles. To focus on active workers, we restrict to workers with at least one contract opened in 2022, leaving 50,598 workers who were active immediately before the release of ChatGPT. Finally, we require each worker to have a selected profile subcategory from which AI exposure can be assigned, leaving 49,610 workers (see Figure \ref{fig:sample-eligibility}).

We use this sample to construct a balanced worker-quarter panel from 2021Q1 through 2026Q1. The unit of observation is a worker $i$ in quarter $q$. Each included worker appears in every quarter; quarters with no contracts are coded as zero. We measure quarterly demand for each worker as the logarithm of the number of contracts the worker is hired for in that quarter:
\begin{equation}
y_{iq}=\log(1+\text{contracts}_{iq}).
\end{equation}

Because the panel is balanced, workers with minimal or no activity after ChatGPT appear as demand declines rather than market exit. A caveat is that we cannot observe workers who closed their accounts entirely before the snapshot, so our estimates should be interpreted as applying to workers whose profiles remained viewable on the platform as of March 2026.

\begin{figure}[H]
\centering
\caption{Distribution of Contracts over Worker-Quarters}
\label{fig:contract-distribution}
\vspace{0.35em}
\includegraphics[width=0.50\textwidth]{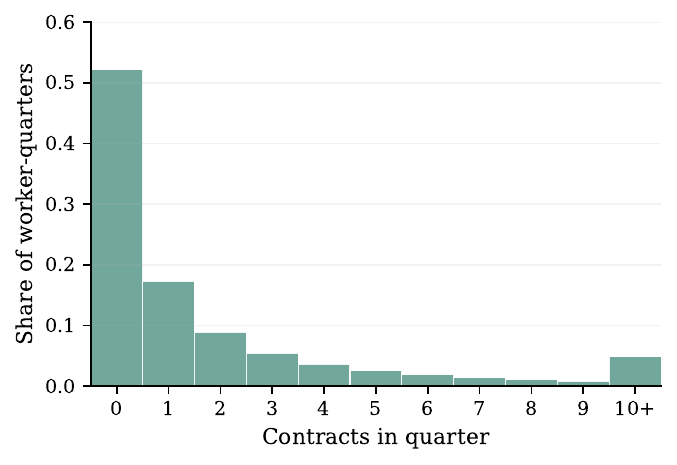}
\vspace{0.25em}
\begin{flushleft}

\end{flushleft}
\end{figure}

Table \ref{tab:sample} summarizes the panel. The final panel contains 49,610 workers observed over 21 quarters, for a total of 1,041,810 worker-quarter observations. The sample contains 2.26 million contracts. Demand is skewed: The mean worker-quarter has 2.17 contracts, while the median worker-quarter has no contracts. This skewness, typical of online labor markets, motivates the use of $\log(1+\text{contracts})$ as the demand outcome; Figure \ref{fig:contract-distribution} shows the distribution of contracts over worker-quarters. The table also reports each worker's AI exposure, a measure of how directly the tasks in the worker's job category are automatable using generative AI, constructed in Section \ref{sec:ai-exposure}.

\begin{table}[H]
\centering
\caption{Sample Summary}
\label{tab:sample}
\vspace{0.4em}
\begin{tabular}{lc}
\toprule
Measure & Value \\
\midrule
Workers & 49,610 \\
Worker-quarters & 1,041,810 \\
Quarters & 21 \\
Sample period & 2021Q1--2026Q1 \\
Total contracts & 2,259,060 \\
Mean contracts per worker-quarter & 2.17 \\
Mean $\log(1+\text{contracts})$ & 0.646 \\
Share of worker-quarters with zero contracts & 52.2\% \\
Mean AI exposure score & 0.252 \\
Standard deviation of AI exposure score & 0.189 \\
\bottomrule
\end{tabular}
\vspace{0.35em}
\begin{flushleft}
\footnotesize Notes: AI exposure is reported in raw ``E1" units from \citet{eloundou2024}.
\end{flushleft}
\end{table}

\subsection{Worker Profiles: Human Capital Signals and Price}

\begin{figure}[H]
\centering
\caption{Worker Profile Information}
\label{fig:worker-profile-blocks}
\vspace{1.00em}
\begin{tikzpicture}[
  node distance=0.40cm,
  block/.style={draw, rounded corners=2pt, text width=3.15cm, minimum height=2.05cm, font=\scriptsize, inner sep=5pt},
  topbrace/.style={decorate, decoration={brace, amplitude=6pt}, thick},
  underbrace/.style={decorate, decoration={brace, mirror, amplitude=6pt}, thick}
]
\node[block] (presentation) {
\begin{minipage}[t][1.65cm][t]{3.0cm}
\centering\textbf{Self-Presentation}\par
\vspace{0.25em}
\raggedright
Title\\
Description\\
Skill Tags\\
Location
\end{minipage}
};
\node[block, right=of presentation] (credentials) {
\begin{minipage}[t][1.65cm][t]{3.0cm}
\centering\textbf{Credentials}\par
\vspace{0.25em}
\raggedright
Education\\
Employment\\
Portfolio
\end{minipage}
};
\node[block, right=of credentials] (reputation) {
\begin{minipage}[t][1.65cm][t]{3.0cm}
\centering\textbf{Reputation}\par
\vspace{0.25em}
\raggedright
Contracts\\
Feedback\\
Ratings\\
Badges
\end{minipage}
};
\node[block, right=of reputation] (wage) {
\begin{minipage}[t][1.65cm][t]{3.0cm}
\centering\textbf{Price}\par
\vspace{0.25em}
\raggedright
Hourly Rate
\end{minipage}
};

\draw[topbrace]
  ([xshift=-0.08cm,yshift=0.42cm]presentation.north west) --
  node[above=0.18cm, font=\footnotesize] {Worker Profile Blocks}
  ([xshift=0.08cm,yshift=0.42cm]wage.north east);
\draw[underbrace]
  ([xshift=-0.08cm,yshift=-0.50cm]presentation.south west) --
  node[below=0.18cm, font=\footnotesize] {Human Capital Signals}
  ([xshift=0.08cm,yshift=-0.50cm]reputation.south east);
\end{tikzpicture}
\vspace{0.85em}
\end{figure}
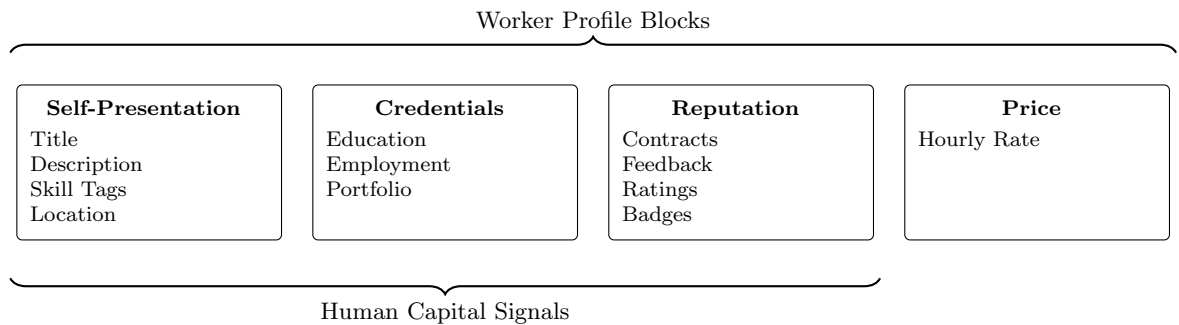

We organize observable worker information into four blocks: three related to signals of the worker's human capital signals plus their posted hourly price. This grouping of human capital signals separates the main sources of pre-hire information: How the worker chooses to present their ability, externally validated education and experience, and prior clients' evaluations from within the market itself. Figure \ref{fig:worker-profile-blocks} shows the main concepts included in each block, and Table \ref{tab:feature-blocks} provides a more detailed list of the variables and describes how they enter the prediction model.

The {\it self-presentation} block includes the worker's chosen title (e.g., ``Experienced iOS Developer"), written profile description, and skill tags. These fields tell clients the worker's stated expertise and emphasized platform-defined skills. Examples of skill tags are provided in Table \ref{tab:skill-tag-examples} in Appendix \ref{app:skill-tags}. The block also includes title length, description length, skill count, and country indicators.

The {\it credentials} block records the worker's accumulated experience and training: Degrees obtained, fields of study, institutions, employment titles and descriptions, and portfolio project titles and descriptions. Although these fields are also reported through the profile, they convey a different kind of information than self-presentation, since they reflect external validation of the worker's human capital. This block is also more persistent than self-presentation -- for example, changing formal education is much more costly for the worker than changing their skill tags.

The {\it reputation} block contains information generated by prior platform activity, including prior contract titles, written feedback comments from past clients, platform certification badges (e.g., ``Top Rated", average prior feedback scores, and the share of prior scored contracts with high feedback. Unlike self-presentation and credentials, reputation information is generated entirely within the market through past transactions. 

The {\it price} block contains the posted hourly rate and its log transform. Because our focus is on the predictive power of price with respect to demand rather than interpreting slopes, multi-collinearity from including both price and its log transform does not undermine our empirical design.

Nearly all workers have profile titles and descriptions, and all workers have skill tags and posted hourly rates. Credentials are also common: 47,976 workers report education (96.7\%), 48,508 report employment history (97.8\%), and 41,066 include at least one project in their portfolio (82.8\%). Reputation is measured from work histories. All workers in the sample have prior platform activity by construction; 48,710 have at least one scored contract (98.2\%), and 47,109 have at least one written feedback comment (95.0\%).

\begin{table}[H]
\centering
\caption{Variables Used Within Each Profile Block}
\label{tab:feature-blocks}
\vspace{0.4em}
\scriptsize
\setlength{\tabcolsep}{4pt}
\renewcommand{\arraystretch}{0.92}
\begin{tabular}{p{0.20\textwidth}p{0.42\textwidth}p{0.20\textwidth}}
\toprule
Component & Variable & Format \\
\midrule
\multirow{8}{*}{Self-Presentation} & Profile Title & Text \\
 & Profile Description & Text \\
 & Skill Tags & Text \\
 & Title Length & Numeric \\
 & Description Length & Numeric \\
 & Skill Count & Numeric \\
 & Country & Categorical \\
 & Availability: Maximum Weekly Hours & Numeric \\
\midrule
\multirow{7}{*}{Credentials} & Education: Degree & Text \\
 & Education: Area of Study & Text \\
 & Education: Institution & Text \\
 & Employment: Job Title & Text \\
 & Employment: Description & Text \\
 & Portfolio: Project Title & Text \\
 & Portfolio: Project Description & Text \\
\midrule
\multirow{9}{*}{Reputation} & Prior Contract Titles & Text \\
 & Prior Feedback Comments & Text \\
 & Top Rated Badge & Binary indicator \\
 & High-Potential Badge & Binary indicator \\
 & Eligible Badge & Binary indicator \\
 & Prior Total Jobs & Numeric \\
 & Prior Total Feedback & Numeric \\
 & Average Prior Feedback Score & Numeric \\
 & Prior High-Score Share & Numeric \\
\midrule
\multirow{2}{*}{Price} & Posted Hourly Rate & Numeric \\
 & Log Posted Hourly Rate & Numeric \\
\bottomrule
\end{tabular}
\vspace{0.35em}
\begin{flushleft}

\end{flushleft}
\end{table}

A key limitation is that we observe worker profiles as a single snapshot from March 2026, so our observables do not exactly replicate the profile pages clients saw at the time of each hiring decision. Nevertheless, the snapshot remains informative. The credential block is unlikely to vary much over time: Educational background, prior employment, portfolio history, and broad occupational positioning are persistent features of a worker's human capital and costly to change. Workers may update skill tags, titles, or descriptions, but are unlikely to switch job categories entirely (e.g., from ``Accountant" to ``Data Engineer"). The reputation variables are an exception: Because they are constructed dynamically from contracts and feedback preceding each quarter, they closely reflect the information available to clients at the time of hiring. 

Finally, although the posted hourly rate is also observed only as of March 2026, prices appear stable over time, and cross-sectional variation in prices is much larger than within-worker temporal variation (see Appendix \ref{app:price-variation}). The March 2026 posted hourly rate is therefore a reasonable approximation of a worker's relative position within the price distribution during the study period. Lastly, we also do not observe workers' cover letters or the order in which workers appear in search results or recommendations, so our model does not include the entire information set available to clients at the time of hiring.

\subsection{AI Exposure}
\label{sec:ai-exposure}

Our continuous treatment variable is each worker's AI exposure, constructed using the occupational AI exposure scores from \citet{eloundou2024}. We use the ``E1" exposure scores from \citet{eloundou2024}, which measure direct LLM exposure: The share of an occupation's tasks for which access to an LLM alone could reduce completion time by at least half while maintaining equivalent quality. These scores vary from 0 to 1, with 1 indicating maximum exposure. Task-based exposure measures are commonly used to study the impact of AI on labor markets \citep{felten2021,acemoglu2022ai,eloundou2024}.

We assign exposure to workers based on their primary job subcategory, defined as the first subcategory listed on the worker's profile. The sample spans 12 major job categories and 102 job subcategories. Examples of subcategories include ``Photography'' (exposure 0.02), ``Accounting'' (exposure 0.20), and ``Legal Translation'' (exposure 0.80). Because the platform's job subcategories differ from the occupations used in \citet{eloundou2024}, we use an LLM to map platform subcategories to occupational categories by semantic similarity -- see Appendix \ref{app:exposure-mapping} for the full list.

In the sample, the mean raw E1 exposure score is 0.252 and the standard deviation is 0.189. The median worker has a raw score of 0.219. Table \ref{tab:upwork-onet-mapping} in Appendix \ref{app:exposure-mapping} shows raw E1 exposure for the mapped job subcategories. Because the most exposed subcategories in the sample (translation services) have a raw score of 0.8, we rescale exposure by this maximum in our main regressions, so that a scaled exposure of 1 corresponds to the most exposed job category and 0 to an unexposed category. This rescaling is for interpretation in the regression tables only.

A caveat is that the worker's selected subcategory is based on the March 2026 snapshot, so exposure is assigned from a post-shock profile field. If workers systematically switched subcategories in response to ChatGPT, exposure would be measured with error. However, we expect this concern to be limited because the worker's selected subcategory reflects their broad occupational positioning, which is unlikely to change dramatically after ChatGPT. Moreover, if workers in heavily exposed categories moved into less exposed ones after the release of ChatGPT, this would bias exposure downward and attenuate our estimate of the treatment effect, making our estimates conservative. It is also unlikely that workers moved toward more AI-exposed job categories, because prior work has shown demand in high-exposure categories was more likely to fall \citep{hui2024, demirci2025}, which we confirm as well.

\section{Empirical Strategy}
\label{sec:strategy}

\begin{figure}[H]
\centering
\caption{Empirical Analysis Pipeline}
\label{fig:analysis-pipeline}
\vspace{1.00em}
\begin{tikzpicture}[
  node distance=0.45cm,
  stepbox/.style={draw, rounded corners=2pt, align=center, text width=2.65cm, minimum height=1.05cm, font=\footnotesize, inner sep=5pt},
  pipelinearrow/.style={-{Stealth[length=2mm]}, thick}
]
\node[stepbox] (embed) {Embed Profile\\Blocks};
\node[stepbox, right=of embed] (predict) {Predict Quarterly\\Labor Demand};
\node[stepbox, right=of predict] (decompose) {Compute Profile \\ Block Importance};
\node[stepbox, right=of decompose] (did) {Estimate\\Exposure DiD};

\node[font=\scriptsize, below=0.08cm of embed] {(\S\hyperref[sec:text-embeddings]{\textcolor{blue}{\ref*{sec:text-embeddings}}})};
\node[font=\scriptsize, below=0.08cm of predict] {(\S\hyperref[sec:predicting-demand]{\textcolor{blue}{\ref*{sec:predicting-demand}}})};
\node[font=\scriptsize, below=0.08cm of decompose] {(\S\hyperref[sec:shapley]{\textcolor{blue}{\ref*{sec:shapley}}})};
\node[font=\scriptsize, below=0.08cm of did] {(\S\hyperref[sec:event-study]{\textcolor{blue}{\ref*{sec:event-study}--\ref*{sec:pooled-did}}})};

\draw[pipelinearrow] (embed) -- (predict);
\draw[pipelinearrow] (predict) -- (decompose);
\draw[pipelinearrow] (decompose) -- (did);
\end{tikzpicture}
\vspace{0.90em}
\begin{flushleft}
\end{flushleft}
\vspace{0.45em}
\end{figure}
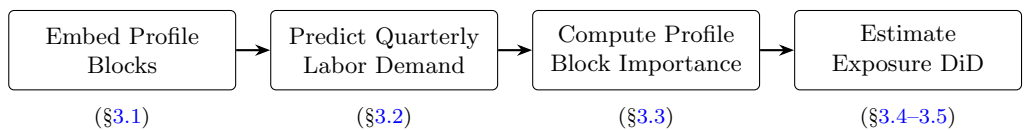

Figure \ref{fig:analysis-pipeline} summarizes the four steps of the empirical strategy. First, we convert worker-profile text into embeddings and combine these with structured profile variables (Section \ref{sec:text-embeddings}). Second, we estimate quarter-specific demand models that map the four worker-profile blocks to demand (i.e., log quarterly contracts) (Section \ref{sec:predicting-demand}). Third, we compute each block's importance for predicted demand using Shapley values (Section \ref{sec:shapley}). Finally, we use these block-importance measures as outcomes in event-study and difference-in-differences specifications by continuous AI exposure (Sections \ref{sec:event-study} and \ref{sec:pooled-did}).

\subsection{Text Embeddings}
\label{sec:text-embeddings}
We begin by converting the text fields in the worker-profile blocks into numeric inputs. Embeddings convert text into numeric vectors \citep{mikolov2013}, allowing us to use profile text in a prediction model without hand-coding specific skills, degrees, or institutions. The vector coordinates are not economically interpretable, but texts with related meanings tend to have similar vectors.

The main specification uses \texttt{all-MiniLM-L6-v2}, a compact model that produces 384-dimensional text embeddings \citep{wang2020minilm}. We use this model because it is widely used and well suited to short and medium-length profile text. As a robustness check, Appendix \ref{app:alter_emb_robustness} reports results using Snowflake Arctic Embed 2.0-M embeddings truncated to 256 dimensions \citep{yu2024arcticembed}. This check addresses both sensitivity to the embedding model and sensitivity to MiniLM's shorter context window, since Arctic can encode substantially longer profile text before truncation.

We construct embeddings for each human capital block. For self-presentation, we concatenate the worker's profile title, profile description, and ordered skill tags, and then embed the combined text. For credentials, we concatenate education, employment, and portfolio project fields, and embed that combined text. For reputation, we construct a quarter-specific text representation for each worker using prior contract titles and client feedback from contracts that began before the quarter. Price is not embedded because the posted hourly rate enters the prediction model numerically. Embeddings are unit-normalized. Missing fields are represented by an empty string before embedding. 

\begin{figure}[H]
\centering
\caption{2D Projection of Self-Presentation Embeddings}
\label{fig:embedding-category-projection}
\vspace{0.35em}
\includegraphics[width=\textwidth]{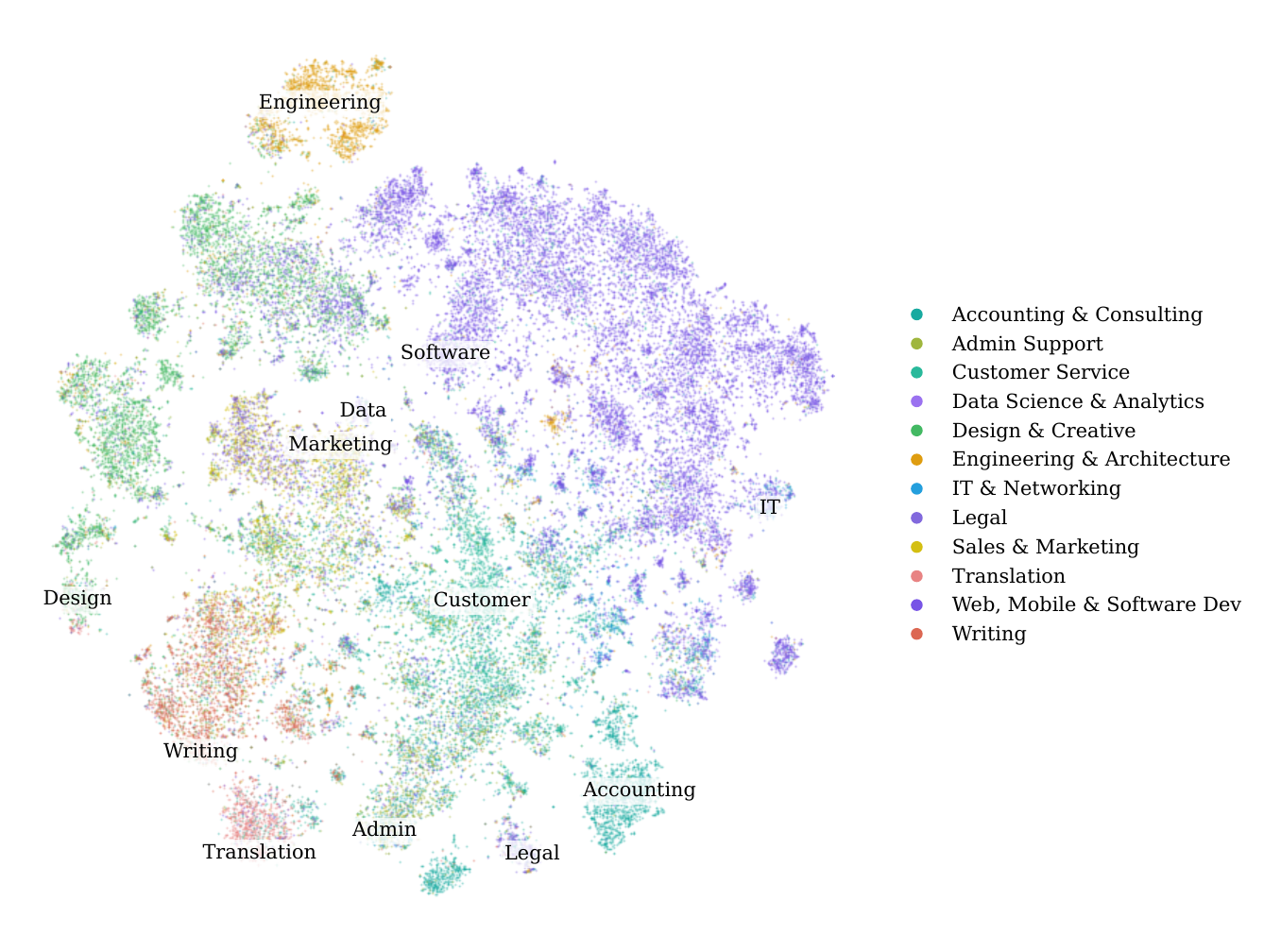}
\begin{flushleft}
\footnotesize Notes: Each point is a worker, represented by the embedding of their profile title, description, and skill tags. The figure uses all 49,610 workers and projects the 384-dimensional MiniLM embeddings into two dimensions using t-SNE \citep{vandermaaten2008}.
\end{flushleft}
\end{figure}

Figure \ref{fig:embedding-category-projection} provides a visualization of the embeddings for the self-presentation block, projected from 384 to 2 dimensions using t-SNE \citep{vandermaaten2008}. We find that a worker's nearest neighbor in the original self-presentation embedding space shares the same broad job category 57.5\% of the time, compared with 18.6\% if workers were paired at random. This informally validates that the self-presentation embeddings are encoding some of the ``meaning" in the worker's self-described occupational positioning and skills, in addition to other information. 

Table \ref{tab:embedding-dimensions} shows the dimensions of each block after combining text embeddings and numeric variables. Although the total number of variables (1,177) is large for OLS, the comparatively larger sample size (49,610) limits overfitting concerns. Moreover, all predictions are generated for out-of-sample workers using a five-fold cross-validation procedure, which further reduces the risk that the main results are influenced by overfitting.
 
\begin{table}[H]
\centering
\caption{Feature Dimensions by Profile Block}
\label{tab:embedding-dimensions}
\vspace{0.35em}
\footnotesize
\setlength{\tabcolsep}{5pt}
\begin{tabular}{lccc}
\toprule
Component & Text Embedding & Numeric Variables & Combined Variables \\
\midrule
Self-presentation & 384 & 16 & 400 \\
Credentials & 384 & 0 & 384 \\
Reputation & 384 & 7 & 391 \\
Price & 0 & 2 & 2 \\
\midrule
Total & 1,152 & 25 & 1,177 \\
\bottomrule
\end{tabular}
\vspace{0.35em}
\begin{flushleft}
\end{flushleft}
\end{table} 

\subsection{Predicting Quarterly Labor Demand}
\label{sec:predicting-demand}
The theoretical object of interest is the market's allocation function: How visible human capital signals and price map into labor demand. Let $\mathcal{K}=\{S,C,R,P\}$ index the four blocks of observable worker information (self-presentation, credentials, reputation, and price), and let $X_{iq}=(X^{k}_{iq})_{k\in\mathcal{K}}$ collect them. We assume the market's allocation of demand across workers in quarter $q$ is described by a function
\begin{equation}
m_q(X_{iq})=\mathbb{E}\left[y_{iq}\mid X_{iq}\right],
\end{equation}
where $y_{iq} = \log(1+\text{contracts}_{iq})$ is the worker's log quarterly demand. 
The allocation function is quarter-specific, which allows us to test whether it changed after the release of ChatGPT. Our focus on the market allocation function is in the spirit of hedonic regressions \citep{rosen1974} and aggregate matching functions used to study labor markets \citep{petrongolo2001}, and allows us to summarize how worker information translates to demand without specifying the underlying search or choice process. 

Accordingly, our empirical goal is not to estimate a structural model, but to measure the importance of the three human capital signals and price in this allocation function. Our main specification approximates the allocation function with a linear model, estimated separately by quarter:
\begin{equation}
y_{iq}
=\beta_{0,q}
+\sum_{k\in\mathcal{K}}\left(\bm{\beta}^{k}_{q}\right)^{\top}X^{k}_{iq}
+\varepsilon_{iq}.
\end{equation}
We use a linear model as the main specification because it is transparent and stable in high dimensions. Appendix \ref{app:alter_emb_robustness} reports robustness checks using ridge regression \citep{hoerl1970} and the highly-flexible XGBoost model \citep{chen2016}; our coefficient estimates remain stable under these alternative prediction methods. Within each quarter, predictions are generated out of fold using five-fold cross-validation, so the fitted demand prediction for worker $i$ is produced by a model estimated without worker $i$.

The first-stage regression is predictive, not causal, following work that uses predictions of economically relevant quantities as inputs to downstream analysis \citep{kleinberg2018}. Accordingly, we do not interpret individual coefficients as causal effects of profile attributes on hiring. Instead, the first-stage prediction model allows us to quantify the importance of each of the four profile blocks, discussed next. 

\subsection{Profile Block Importance}
\label{sec:shapley}

The prediction model predicts demand but does not identify which profile blocks are important for a particular worker's demand. One intuitive approach would be to measure how much the regression model's $R^2$ decreases when one block is omitted \citep{Lei2018}. However, when blocks are correlated, this ``leave-one-out" analysis can understate the importance of a block, and the resulting importance scores are not guaranteed to sum to the model's total explanatory power \citep{owen2014,covert2020}. We instead compute Shapley regression values, which are widely used to measure variable importance in machine learning models \citep{lipovetsky2001analysis,lundberg2017}. Our block-level implementation follows the tradition of group-level Shapley decompositions of model fit \citep{huettner2012}.

Shapley values treat the blocks symmetrically by averaging the marginal contribution of each block across all possible combinations in which the blocks could enter the model. Specifically, let $\hat{y}_{iq}(S)$ denote the out-of-fold prediction for worker $i$ in quarter $q$ from a model that uses only the subset of blocks $S\subseteq\mathcal{K}$, estimated on cross-validation folds that exclude worker $i$. The Shapley value for block $k$ is then
\begin{equation}
\phi^k_{iq}
=
\sum_{S\subseteq \mathcal{K}\setminus\{k\}}
\frac{|S|!(|\mathcal{K}|-|S|-1)!}{|\mathcal{K}|!}
\left[\hat{y}_{iq}(S\cup\{k\})-\hat{y}_{iq}(S)\right].
\end{equation}
With four blocks, we compute these values by fitting all $2^4=16$ possible models based on subsets of blocks. By construction, the four Shapley values decompose predicted demand exactly, relative to the mean demand, $\bar{y}_q$:
\begin{equation}
\hat{y}_{iq}
=
\bar{y}_q+\sum_{k\in\mathcal{K}}\phi^{k}_{iq}.
\label{eq:adding-up}
\end{equation}

The Shapley value $\phi_{iq}^k$ thus measures the {\it observation-level} importance of block $k$ for worker $i$'s demand in quarter $q$ -- that is, the extent to which the block shifts the predicted allocation of demand for that specific worker-quarter. The Shapley values are computed based on out-of-sample demand predictions within the cross-validation procedure, to minimize distortion of the block-importance measures due to overfitting.

Because the self-presentation, credentials, and price blocks are fixed at their snapshot values, movements in their Shapley values over time reflect changes in the allocation function rather than changes in worker attributes. The exception is the reputation block, which is time-varying. However, this does not undermine the decomposition, because reputation is predetermined within quarter $q$, so its Shapley value still measures how strongly the allocation function uses the reputation information available to clients at that time.

\subsection{Event-Study Specification}
\label{sec:event-study}

We adopt a difference-in-differences framework using the release of ChatGPT as the focal shock, repeating the analysis for each of the four blocks $k \in \{P, C, R, W\}$ separately. Let
\begin{equation}
\Delta_k(a)
\equiv
\mathbb{E}\!\left[\phi^k_{iq}\mid E_i=a,\,\text{Post}_q=1\right]
-
\mathbb{E}\!\left[\phi^k_{iq}\mid E_i=a,\,\text{Post}_q=0\right]
\end{equation}
denote the pre-to-post change in the block-$k$ Shapley value at AI exposure level $E_i = a$, with the expectation taken over all worker-quarters $(i,q)$.  For two exposure levels $a_1>a_0$, the estimand for block $k$ is the difference in these changes:
\begin{equation}
\theta_k(a_1,a_0)=\Delta_k(a_1)-\Delta_k(a_0).
\end{equation}
This estimand captures whether the importance of information block $k$ in the allocation of demand changed more after ChatGPT for workers in more AI-exposed categories than for workers in less exposed categories. 
For each outcome $Y_{iq}\in\{y_{iq},\phi^{S}_{iq},\phi^{C}_{iq},\phi^{R}_{iq},\phi^{P}_{iq}\}$, where $y_{iq}$ is the demand of worker $i$ during quarter $q$, we estimate the event-study
\begin{equation}
\begin{aligned}
Y_{iq}
=\beta_0
+\rho E_i
+\sum_{\tau\neq 2022Q3}\lambda_{\tau}\mathbf{1}\{q=\tau\}
+\sum_{\tau\neq 2022Q3}\delta_{\tau}
\left(E_i\times\mathbf{1}\{q=\tau\}\right)
+u_{iq},
\end{aligned}
\label{eq:event-study}
\end{equation}
ChatGPT was released on November 30, 2022, late in 2022Q4. We therefore use 2022Q3 as the omitted quarter, so the reference period is fully pre-release, and treat 2022Q4 as the first post-release quarter. Because only December 2022 falls after the release, the 2022Q4 coefficient is diluted and likely conservative. 

The identifying assumption is that, absent the availability of AI technology, workers in more and less AI-exposed categories would have followed parallel relative trends in the importance of each profile block. We examine this assumption visually using pre-treatment coefficients in the event-study plots, and formally using joint Wald tests of the pre-treatment interactions (Appendix \ref{app:pretrends}). Because the statistical significance of pre-trend tests does not by itself indicate whether deviations are economically meaningful, we follow \citet{roth2022} and report the magnitudes of the pre-period coefficients alongside the tests. 

\subsection{Difference-in-Differences}
\label{sec:pooled-did}

We summarize the post-AI change using a pooled difference-in-differences specification:
\begin{equation}
Y_{iq}
=\beta_0+\rho E_i+\lambda \text{Post}_q+\theta(E_i\times \text{Post}_q)+u_{iq},
\label{eq:pooled-did}
\end{equation}
where $\text{Post}_q = 1$ beginning in $2022Q4$. We report $\theta$ as the contrast between the most and least exposed categories, comparing a category with a scaled exposure of 1 (the maximum among workers in our panel) to a category with an exposure of 0; this is the linear continuous-treatment analogue of $\theta_k(a_1,a_0)$ above. The coefficient is in log-demand units, matching the first-stage predictions. Because treatment effects may vary with the length of exposure \citep{callaway2021multiple}, we also estimate a version that splits $\text{Post}_q$ into an early window (2022Q4--2025Q1) and a late window covering the final four sample quarters (2025Q2--2026Q1), with separate exposure interactions. This multiple post-period specification is also motivated by evidence that labor-market effects of generative AI can evolve as adoption diffuses \citep{demirci2025}. 
 
Because the Shapley values are obtained from a first-stage prediction model, treating them as observed variables may understate uncertainty. We therefore use a worker-level bootstrap to construct standard errors. Each bootstrap repetition resamples workers and repeats the second, third and fourth steps of the empirical analysis (as shown in Figure \ref{fig:analysis-pipeline}). We use 100 repetitions to compute bootstrap $p$-values. Our sample is close to exhaustive over the relevant market population: it covers all subcategories and searchable workers on the platform, and AI exposure depends deterministically on job subcategories. This motivates our use of worker-level resampling for our primary inference, following the design-based perspective in \citet{abadie2023}. Because the identifying variation in exposure is nevertheless at the subcategory level, Appendix \ref{app:permutation} reports a placebo test that permutes exposure scores across subcategories.

\section{Results}
\label{sec:results}

\subsection{AI Exposure and Labor Demand}
\label{sec:demand-benchmark}

\begin{figure}[H]
\centering
\caption{Observed Demand Event Study}
\label{fig:demand-event-study}
\vspace{0.35em}
\includegraphics[width=0.50\textwidth]{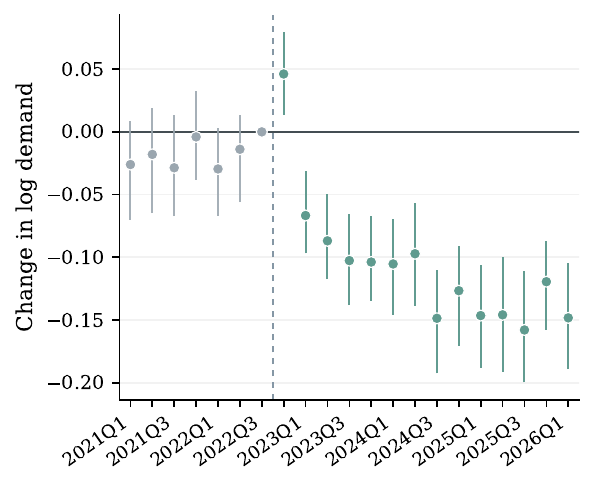}
\vspace{0.25em}
\begin{flushleft}
\footnotesize Notes: The dashed vertical line marks the first post-AI quarter. Intervals are percentile bootstrap intervals from the worker-level bootstrap.
\end{flushleft}
\end{figure}

To establish a baseline, we begin with the demand response itself, where the left-hand side of Equations \ref{eq:event-study} and \ref{eq:pooled-did} is the worker's demand, $y_{iq}$, measured as the logarithm of the number of contracts they are hired for. Figure \ref{fig:demand-event-study} plots the event-study coefficients. Demand falls for workers in more AI-exposed categories after ChatGPT: The pooled estimate, reported in column (1) of Table \ref{tab:pooled-did-main}, is $-0.0725$ (comparing the most exposed worker with an unexposed worker). This implies that after ChatGPT, contract volume for workers in the most exposed job category fell by approximately 7.0\% relative to workers in an unexposed job category.

This result is not novel -- it aligns with the general demand decline documented by \citet{hui2024} and \citet{demirci2025}. We use this result to anchor the block importance estimates: Because the Shapley values sum to predicted demand (Equation~\ref{eq:adding-up}), the results that follow must collectively account for a decline of roughly this size. The rest of this section asks how that demand decline is distributed across the three human capital blocks and price.

\subsection{Block Importance: Human Capital Signals and Price}

\begin{figure}[H]
\centering
\caption{Block Importance by AI Exposure}
\label{fig:bootstrap-shapley-event-study}
\vspace{0.35em}
\includegraphics[width=0.95\textwidth]{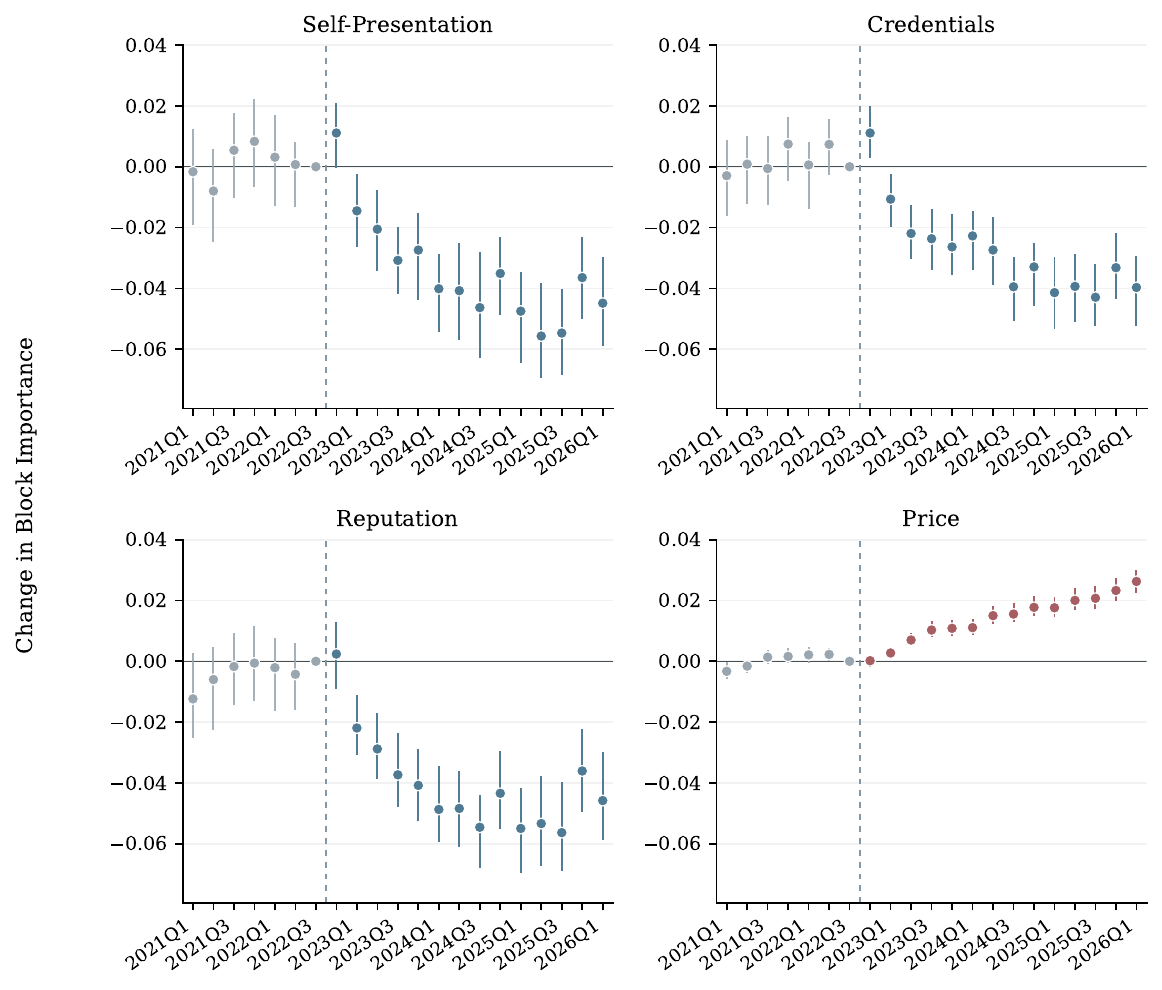}
\vspace{0.25em}
\begin{flushleft}
\footnotesize Notes: The figure rescales the event-study estimates to show the effect of moving from an unexposed category (scaled exposure 0) to the most exposed category (scaled exposure 1) on the Shapley value of each block, relative to 2022Q3. The dashed vertical line marks the first post-AI quarter. Intervals are percentile bootstrap intervals from the worker-level bootstrap.
\end{flushleft}
\end{figure}

We now address our main research question of whether generative AI has changed the importance of workers' human capital signals and price with respect to the demand. Figure \ref{fig:bootstrap-shapley-event-study} plots the event-study estimates for the four profile blocks. The importance of all three human capital signals declines significantly after the release of ChatGPT in more AI-exposed categories. By contrast, the importance of price increases. 

The human capital signals show little differential movement by AI exposure prior to the release of ChatGPT. Joint Wald tests of the pre-treatment coefficients fail to reject parallel pre-trends, with bootstrap $p$-values of 0.28 for self-presentation, 0.28 for credentials, and 0.61 for reputation. The price component shows statistically significant pre-period movement (joint $p=0.01$), but the deviations are small relative to the treatment effect (approximately one-seventh). Additional details regarding the pre-trends are in Appendix \ref{app:pretrends}. 

\begin{table}[H]
\centering
\small
\caption{Difference-in-Differences Estimates for Demand and Block Importance by AI Exposure}
\label{tab:pooled-did-main}
\vspace{0.4em}
\begin{tabular}{lccccc}
\toprule
 & Observed Demand & \multicolumn{4}{c}{Profile Block Importance} \\
\cmidrule(lr){2-2}\cmidrule(lr){3-6}
 & & Self-Presentation & Credentials & Reputation & Price \\
 & (1) & (2) & (3) & (4) & (5) \\
\midrule
\multicolumn{6}{l}{\textit{Baseline}} \\
$\text{Exposure} \times \text{Post}$ & $-0.0725\sym{***}$ & $-0.0285\sym{***}$ & $-0.0238\sym{***}$ & $-0.0294\sym{***}$ & $0.0111\sym{***}$ \\
 & $(0.0117)$ & $(0.0041)$ & $(0.0027)$ & $(0.0050)$ & $(0.0010)$ \\
\addlinespace
\multicolumn{6}{l}{\textit{Late Post-Period}} \\
$\text{Exposure} \times \text{LatePost}$ & $-0.1006\sym{***}$ & $-0.0393\sym{***}$ & $-0.0325\sym{***}$ & $-0.0352\sym{***}$ & $0.0178\sym{***}$ \\
 & $(0.0146)$ & $(0.0052)$ & $(0.0036)$ & $(0.0061)$ & $(0.0014)$ \\
\midrule
Observations & 1,041,810 & 1,041,810 & 1,041,810 & 1,041,810 & 1,041,810 \\
Workers & 49,610 & 49,610 & 49,610 & 49,610 & 49,610 \\
\bottomrule
\end{tabular}
\vspace{0.35em}
\begin{flushleft}
\footnotesize Notes: Column (1) uses $\log(1+\text{contracts})$; columns (2)--(5) use Shapley values for predicted log demand. Worker-level bootstrap standard errors are in parentheses. Post begins in 2022Q4; the late-post specification includes a separate unreported interaction for 2022Q4--2025Q1. \sym{*} $p<0.10$, \sym{**} $p<0.05$, \sym{***} $p<0.01$.
\end{flushleft}
\end{table}

Table \ref{tab:pooled-did-main} reports the main estimates: Column (1) shows the demand benchmark from Section \ref{sec:demand-benchmark}, and columns (2)--(5) report the block importance results. Because exposure is scaled from 0 to 1, each coefficient compares the post-ChatGPT change for workers in the most AI-exposed categories (e.g., Translation) to the change for workers in unexposed categories (e.g., Audio \& Music Production). Expressed as percentage changes, the importance of self-presentation, credentials, and reputation falls by approximately 2.8\%, 2.4\%, and 2.9\%, respectively. Taken together, the importance of the human capital signals falls by 7.8\%. Price moves in the opposite direction, rising by 1.1\%.

The effects also strengthen over time. The second specification in Table \ref{tab:pooled-did-main} reports the exposure interaction for the final four sample quarters (2025Q2--2026Q1). Over this late window, demand falls by 9.6\%, the combined importance of human capital signals falls by 10.1\%, and price importance rises by 1.8\%. The market's reallocation of demand away from human capital signals and toward price is therefore not confined to the immediate ChatGPT release, but continues to widen more than three years after the shock. 

By construction, the four Shapley values sum to predicted demand relative to the quarter mean (Equation \ref{eq:adding-up}). Therefore, because realized demand falls in more exposed categories after ChatGPT (Section \ref{sec:demand-benchmark}), the four blocks must jointly decline by a similar magnitude. The aggregate Shapley-value decline in Table \ref{tab:pooled-did-main} thus follows mechanically from the demand decline in column (1) of that table. However, the result of interest is how this decline is distributed across the four profile blocks: The decline could have been concentrated in any single block, yet it is spread almost evenly across the three human capital blocks. Most importantly, the importance of price rises, despite the mechanical decline, suggesting a substantive shift toward price in the market's allocation function for demand.

Appendix \ref{app:robustness} reports robustness tests on two alternative prediction models, ridge regression and XGBoost, as well as a specification that allows the pre-to-post change to differ by workers' posted prices. The decline in the importance of human capital signals holds throughout.

The decline in self-presentation is consistent with \citet{cowgill2026does}, \citet{cui2025signaling}, and \citet{galdin2025making}, who find that worker self-presentation (e.g., through written description) loses informational value after the availability of generative AI tools, due to cheap talk \citep{crawford1982}. However, our results show that clients may also place less value on {\it verifiable} human capital signals, such as education and employment history, suggesting a broader mechanism than cheap talk alone. We investigate this mechanism next.

\subsection{Mechanism: Commoditization of Labor?}
\label{sec:mechanism}

Experimental evidence shows that generative AI may weaken the association between human capital and productivity, in particular by raising the performance of lower-skilled workers more than that of higher-skilled workers, compressing variation in output quality \citep{noy2023,peng2023,dellacqua2023,brynjolfsson2025}. This suggests a mechanism to explain our results in Figure \ref{fig:bootstrap-shapley-event-study} and Table \ref{tab:pooled-did-main}: If demand-side clients anticipate that AI compresses workers' output quality, they may place less weight on signals of human capital in hiring decisions, and demand may reallocate along the remaining dimension that still differentiates workers: price. A similar mechanism is formalized in the theoretical model of \citet{fukui2026}, who refer to this shift as the ``commoditization of labor''.

Concretely, labor commoditization has two testable implications in our setting. The first is that the ``demand premium" commanded by workers with strong human capital should weaken in more AI-exposed categories, because clients expect an improvement in the output of workers with low human capital. The second is that demand should reallocate toward lower-priced workers more strongly in high-exposure categories, for the same reason. Our analysis offers evidence of both. 

To test the first implication, we classify workers into the top and bottom halves of pre-AI human capital strength. Specifically, define
\begin{equation}
H_i
=
\frac{1}{|\mathcal{Q}_{pre}|}
\sum_{q\in\mathcal{Q}_{pre}}
\frac{1}{3}
\left(
\phi^S_{iq}+\phi^C_{iq}+\phi^R_{iq}
\right),
\label{eq:hc-signal-score}
\end{equation}
where $\mathcal{Q}_{pre}$ denotes pre-ChatGPT quarters. Because each Shapley value $\phi^k_{iq}$ measures how much block $k$ moves worker $i$'s predicted demand above or below the quarter mean, $H_i$ represents the pre-AI demand premium attributable to the worker's observable human capital signals. Workers above the median of $H_i$ are thus assigned to the ``high" human capital group. We then estimate a triple-difference model that tests whether the demand premium attached to high human capital signals compresses more after ChatGPT in more AI-exposed categories.

For the second test, we repeat the comparison described above for workers in the top and bottom halves of posted price. In our sample, higher-priced workers also see higher demand before ChatGPT. The price test therefore asks whether this demand gap shifts toward lower-priced workers after ChatGPT. An important source of ambiguity in our analysis thus far is that prices themselves may signal quality \citep{milgrom1986}, in which case the rising importance of price in our main results in Table \ref{tab:pooled-did-main} could reflect clients using price as a substitute quality signal, rather than increased commoditization. The key question is therefore whether demand shifts toward lower-priced workers, as the commoditization interpretation would predict.

\begin{figure}[H]
\centering
\caption{Raw Demand Levels by Human Capital and Price Groups}
\label{fig:mechanism-raw-levels}
\vspace{0.35em}
\includegraphics[width=0.90\textwidth]{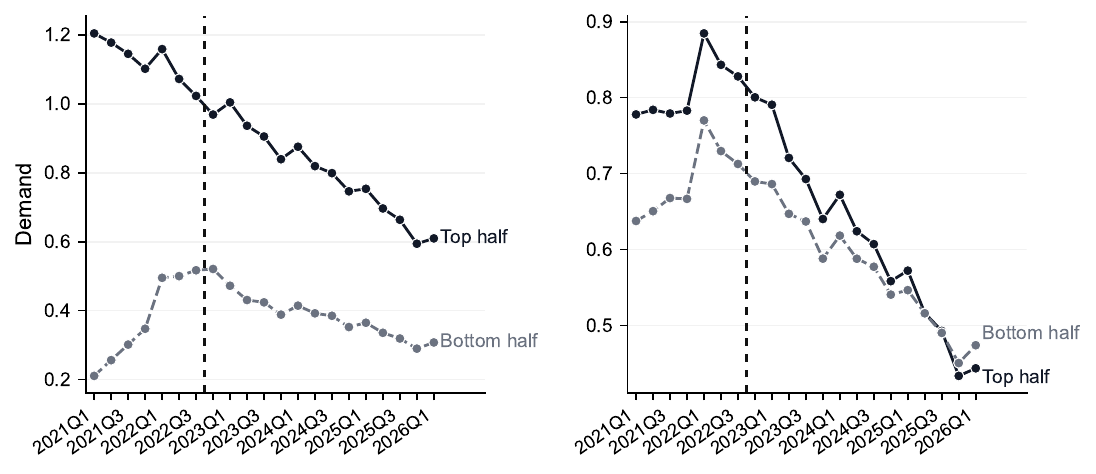}
\vspace{0.25em}
\begin{flushleft}
\footnotesize Notes: The left panel plots mean log quarterly demand for workers in the top and bottom halves of pre-AI human capital signal strength, measured as the average Shapley importance value across three human capital blocks (Equation \ref{eq:hc-signal-score}). The right panel plots the corresponding raw demand levels for workers in the top and bottom halves of posted price. The dashed vertical line marks the first post-AI quarter.
\end{flushleft}
\end{figure}

As an initial visualization, Figure \ref{fig:mechanism-raw-levels} plots raw demand levels, pooled across all job categories. Both plots depict a narrowing of the demand gap over time. These unconditional trends provide useful intuition, but they do not show whether the gaps narrow {\it more} in categories with greater AI exposure, which would be necessary for a causal interpretation. The triple-difference tests below therefore ask whether the worker-level demand gap narrows more after ChatGPT in more AI-exposed categories.\footnote{An alternative approach would collapse the data to the job category level and estimate a simpler difference-in-differences model using the demand gap within each category as the outcome. We keep the analysis at the worker level because the main outcomes of interest, the Shapley values for each profile block, are measured at the worker level.}

\begin{figure}[H]
\centering
\caption{Demand Gap Compression by AI Exposure}
\label{fig:mechanism-ddd-panel}
\vspace{0.35em}
\includegraphics[width=0.90\textwidth]{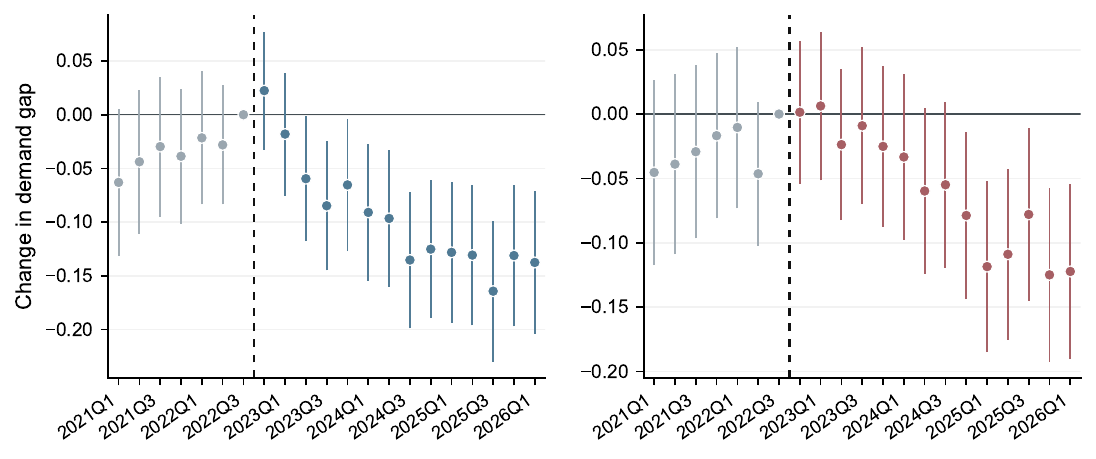}
\vspace{0.25em}
\begin{flushleft}
\footnotesize Notes: The left panel plots the event-study coefficient corresponding to $\text{Top}_i^{\mathrm{HC}} \times E_i$, where $\text{Top}_i^{\mathrm{HC}} = 1$ indicates the worker is in the top half of the human capital strength distribution and $E_i \in [0,1]$ is the worker's AI exposure score. The right panel plots the corresponding coefficient for $\text{Top}_i^{\mathrm{P}} \times E_i$, where $\text{Top}_i^{\mathrm{P}} = 1$ if the worker's posted price is in the top half of all prices. Negative values indicate that the human capital demand gap narrows, or that demand shifts toward lower-priced workers, more in more AI-exposed categories. The dashed vertical lines mark the first post-AI quarter.
\end{flushleft}
\end{figure}

Figure \ref{fig:mechanism-ddd-panel} plots the corresponding event-study estimates. Each coefficient is the quarter-specific interaction between the top-half indicator and AI exposure, measured relative to 2022Q3. For human capital, negative values indicate that the demand gap between workers with high and low human capital is smaller in more AI-exposed categories. For price, negative values indicate that demand shifts toward lower-priced workers more strongly in more AI-exposed categories. Both patterns are consistent with labor commoditization.

To summarize the event-study patterns, we estimate the triple-difference specification
\begin{equation}
\begin{aligned}
Y_{iq}
=&\alpha
+\beta \text{Top}_i^k
+\rho E_i
+\lambda \text{Post}_q
+\psi (\text{Top}_i^k\times E_i) \\
&+\kappa (\text{Top}_i^k\times \text{Post}_q)
+\mu (E_i\times \text{Post}_q) \\
&+\theta_k(\text{Top}_i^k\times E_i\times \text{Post}_q)
+u_{iq},
\end{aligned}
\label{eq:mechanism-ddd}
\end{equation}
where $Y_{iq}=\log(1+\text{contracts}_{iq})$, $k\in\{\mathrm{HC},\mathrm{P}\}$, and $E_i$ is worker $i$'s scaled AI exposure. $\text{Top}_i^{\mathrm{HC}}$ indicates that worker $i$ is in the top half of pre-AI human capital strength, as defined in Equation \ref{eq:hc-signal-score}; $\text{Top}_i^{\mathrm{P}}$ indicates that worker $i$ is in the top half of posted price. In both specifications, the comparison group is the corresponding bottom half, and $\text{Post}_q=1$ beginning in 2022Q4. The coefficient $\theta_k$ measures whether the top-minus-bottom demand gap compresses more after ChatGPT in more AI-exposed categories, with $\theta_k<0$ indicating compression. For example, $\theta_k=-0.01$ means that, after ChatGPT, the top-minus-bottom demand gap narrows by an additional 0.01 log points (about 1\%) when comparing a fully exposed category to an unexposed category.

\begin{table}[H]
\centering
\caption{Triple-Difference Estimates for Demand Gap Compression}
\label{tab:mechanism-pooled}
\vspace{0.4em}
\begin{tabular}{lcc}
\toprule
 & Human Capital & Price \\
 & (1) & (2) \\
\midrule
\multicolumn{3}{l}{\textit{Baseline}} \\
$\text{Top} \times \text{Exposure} \times \text{Post}$ & $-0.0640\sym{***}$ & $-0.0325$ \\
 & $(0.0209)$ & $(0.0225)$ \\
\addlinespace
\multicolumn{3}{l}{\textit{Late Post-Period}} \\
$\text{Top} \times \text{Exposure} \times \text{LatePost}$ & $-0.1089\sym{***}$ & $-0.0818\sym{***}$ \\
 & $(0.0258)$ & $(0.0278)$ \\
\midrule
Observations & 1,041,810 & 1,041,810 \\
Workers & 49,610 & 49,610 \\
\bottomrule
\end{tabular}
\vspace{0.35em}
\begin{flushleft}
\footnotesize Notes: In column (1), $\text{Top}$ corresponds to $\text{Top}_i^{\mathrm{HC}}$, an indicator for workers in the top half of pre-AI human capital Shapley values. In column (2), $\text{Top}$ corresponds to $\text{Top}_i^{\mathrm{P}}$, an indicator for workers in the top half of posted price. Worker-clustered standard errors are in parentheses. Post-period begins in 2022Q4; the late-post specification includes a separate unreported EarlyPost triple interaction for 2022Q4--2025Q1, in addition to LatePost corresponding to 2025Q2--2026Q1. \sym{*} $p<0.10$, \sym{**} $p<0.05$, \sym{***} $p<0.01$.
\end{flushleft}
\end{table}

Figure \ref{fig:mechanism-ddd-panel} and Table \ref{tab:mechanism-pooled} suggest a stronger reallocation of labor demand after ChatGPT in more exposed categories, especially later in the study period. For the human capital specification, the pooled estimate is $-0.064$ log-demand units (equivalent to a 6.2\% decline). For the price specification, the pooled triple-difference estimate is smaller and less precise, at $-0.033$ log-demand units (-3.2\%). As with the block importance results, the effect strengthens over time: In the second specification in Table \ref{tab:mechanism-pooled}, the human-capital demand-gap decline reaches $-0.109$ log-demand units (-10.3\%) in the final four quarters, and the relevant price coefficient reaches $-0.082$ (-7.9\%), and becomes statistically significant at the 5\% level. Taken together, these patterns are consistent with a mechanism in which the market views labor as more substitutable after AI, weakening the association between human capital and demand, and shifting more demand toward lower-priced workers.

We note here that the top and bottom human capital groups are constructed from Shapley values from our demand prediction model, so they are naturally correlated with observed demand; in the pre-period, the worker-level Pearson correlation between $H_i$ and observed log demand is 0.71. The human capital compression result is thus closely related to the more general question of whether ``high-demand" workers see their demand premium decrease post-AI. However, because we focus specifically on strong human capital workers (rather than high demand workers more generally), our test is more relevant for identifying labor commoditization.

A related question is how closely the top-vs-bottom price grouping correlates with the human capital grouping; if the groups are similar, then the right panel of Figure \ref{fig:mechanism-ddd-panel} would be duplicative of the left. However, the price split is very weakly correlated with human capital strength: The $\text{Top}_i^{\mathrm{P}}$ and $\text{Top}_i^{\mathrm{HC}}$ indicators have a Pearson correlation of only 0.070. The price result in column (2) of Table \ref{tab:mechanism-pooled} is therefore not mechanically implied by the human capital result in column (1); it provides separate evidence consistent with labor commoditization. 

The result in column (1) of Table \ref{tab:mechanism-pooled} is related to \citet{hui2024}, who find that top workers in writing-related occupations experience larger demand declines after ChatGPT. We differ in two ways. First, we define workers with high human capital using the pre-AI demand importance of human capital signals, rather than separate proxies for worker quality or prior success. Second, we use continuous AI exposure as treatment over all available job categories, instead of focusing on writing-related tasks only. The more important distinction is that \citet{hui2024} do not investigate demand reallocation toward lower-priced workers (our Table \ref{tab:mechanism-pooled} column (2)), which is central to interpreting the evidence as labor commoditization.

As noted earlier, the results in Table \ref{tab:pooled-did-main}  alone cannot distinguish quality-signaling from commoditization, because both would imply that a worker's price becomes more predictive of their demand. However, the reallocation of demand toward lower-priced workers observed in Figure \ref{fig:mechanism-ddd-panel} and Table \ref{tab:mechanism-pooled} suggests that increased price sensitivity dominates any potential increase in the signaling value of high prices. If clients increasingly interpreted high prices as evidence of quality after ChatGPT, demand should have shifted {\it toward} higher-priced workers in exposed categories. The observed movement in the other direction suggests that increased price sensitivity prevails, although the two effects may still coexist.

\section{Discussion}
\label{sec:discussion}

This paper presents empirical evidence that AI technology may have a commoditizing effect on labor in an online labor market. Prior work has documented that AI may lead to compression in worker output quality \citep{noy2023,dellacqua2023,brynjolfsson2025}, weakening the relationship between human capital and productivity. This paper builds on that literature by also documenting a demand-side response in how clients value workers' human capital, with the human capital premium eroding more in AI-exposed job categories. Because we do not observe demand from individual clients, this market-level response may reflect changes in how individual clients evaluate workers, changes in the composition of active clients, or both. Our results are also consistent with and provide empirical support for the theoretical model of labor commoditization proposed by \citet{fukui2026}.

Labor commoditization has implications for the design and management of online labor markets. Search and recommendation systems trained on historical data may need revision if AI changes the relationship between profile information, posted prices, and hiring fit. Badges and other certification schemes may also become less informative if clients place less weight on conventional human capital signals. 

For workers, the results suggest that incentives to invest in conventional human capital signals may weaken. Workers accumulate these signals deliberately, by pursuing skill certifications, assembling project portfolios, and building their reputation. Our estimates suggest that the demand return to these signals fell in AI-exposed categories and may continue to fall as AI capabilities improve and adoption becomes more widespread. This pattern may be especially consequential for how workers pursue early reputation formation, which prior work has shown can increase future hiring \citep{pallais2014}.

This shift in how demand is allocated may also have consequences for worker welfare. If generative AI has a commoditizing effect on labor, workers with stronger human capital signals may lose part of their competitive advantage in the market. Greater price sensitivity may also intensify price competition in the long run, forcing workers to lower prices and capture less of the surplus they create.

Freelance labor markets may be especially prone to AI-based commoditization because work is often discrete, contract-based, and organized around clear deliverables. At the same time, the patterns we find may emerge in other labor markets; employers in traditional labor markets also screen candidates through resumes, credentials, and references, and human capital is a driver of compensation more generally \citep{neal1995,lazear2009}. A similar labor commoditization effect could emerge in traditional employment, which may alter the returns to different forms of human capital and workers' incentives to invest in them.

{\footnotesize
\bibliographystyle{apalike}
\bibliography{references}
}

\appendix

\section{Robustness}
\label{app:robustness}

\subsection{Alternative Prediction and Embedding Models}
\label{app:alter_emb_robustness}

Our main results are based on a linear regression prediction model for worker demand. We additionally consider two commonly used machine learning methods -- ridge regression \citep{hoerl1970} and XGBoost \citep{chen2016} -- and one alternative embedding model, Snowflake Arctic Embed 2.0-M with embeddings truncated to 256 dimensions \citep{yu2024arcticembed}. 

\begin{figure}[H]
\centering
\caption{Predictive Performance Across Models and Embeddings}
\label{fig:prediction-fit}
\vspace{0.35em}
\includegraphics[width=0.92\textwidth]{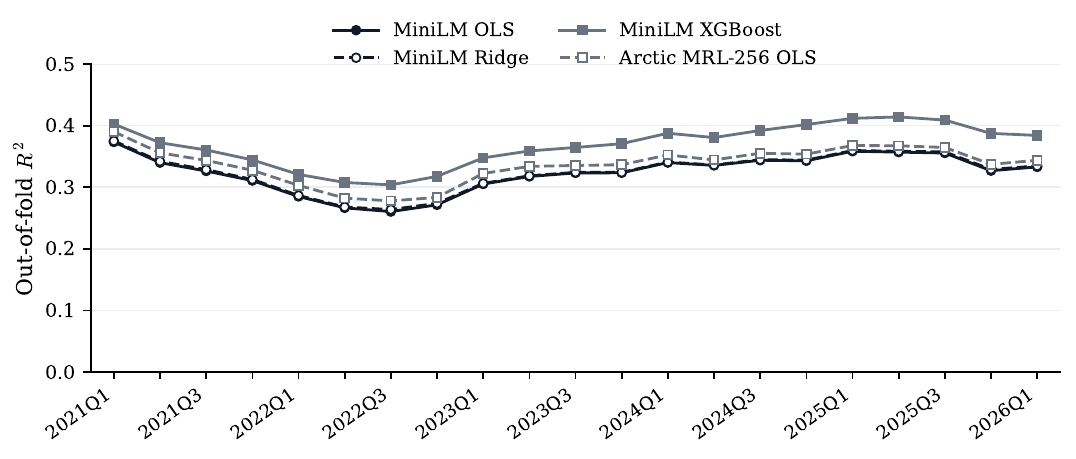}
\vspace{0.25em}
\begin{flushleft}
\footnotesize Notes: The figure plots quarter-level out-of-fold $R^2$ across the 21 sample quarters. All models predict $\log(1+\text{contracts})$ using the worker-profile blocks described in Table \ref{tab:feature-blocks}.
\end{flushleft}
\end{figure}

We first compare predictive performance, reported in Figure \ref{fig:prediction-fit}. In general, predictive accuracy is stable across quarters and comparable across methods. The MiniLM OLS specification on which our main results are based has a median $R^2$ of 0.327 across quarters. Ridge regression performs nearly identically. A nonlinear XGBoost model improves median $R^2$ to 0.372, while the 256-dimensional Arctic Embed 2.0-M embeddings with OLS improve median $R^2$ to 0.344. Across specifications, predictive performance is stable in the post-period (after 2022Q3), suggesting that the observed changes in human capital and price importance are not artifacts of changes in model fit quality. 

The Arctic specification also clarifies two implementation details. First, all text is capped at 2,000 characters before either model encodes it; for reputation histories, this retains the newest prior contract information first. Second, Arctic has a much longer token sequence limit than MiniLM: MiniLM uses its default 256-token limit, while Arctic is run with a 1,536-token limit. During our embedding step, no 2,000-character-capped text exceeds Arctic's 1,536-token limit. Arctic's native embeddings are 768-dimensional, and we use its Matryoshka Representation Learning (MRL) truncation to reduce them to 256 dimensions; this 256-dimensional representation retains nearly all of the full-dimensional retrieval performance \citep{yu2024arcticembed}.

We next examine the sensitivity of the difference-in-differences estimates to the demand-prediction model and choice of text embeddings. Table \ref{tab:robust-pooled} provides a summary. The estimates are stable across both alternative prediction models and under the alternative 256-dimensional Arctic Embed 2.0-M embeddings. In every specification, self-presentation, credentials, and reputation decline while price rises, with sign-based bootstrap $p$-values below 0.01 for all four outcomes. For the three secondary specifications, bootstrap standard errors use an $m$-out-of-$n$ worker bootstrap \citep{bickel1997resampling}: 100 repetitions resample 10\% of workers, with draw deviations scaled by $\sqrt{m/N}$, where $m$ is the bootstrap worker sample size and $N$ is the full worker sample.

\begin{table}[H]
\centering
\caption{Estimates from Alternative Prediction Models}
\label{tab:robust-pooled}
\vspace{0.4em}
\begin{tabular}{lcccc}
\toprule
Specification & Self-Presentation & Credentials & Reputation & Price \\
\midrule
MiniLM OLS & $-0.0285\sym{***}$ & $-0.0238\sym{***}$ & $-0.0294\sym{***}$ & $0.0111\sym{***}$ \\
 & $(0.0041)$ & $(0.0027)$ & $(0.0050)$ & $(0.0010)$ \\
MiniLM Ridge & $-0.0287\sym{***}$ & $-0.0231\sym{***}$ & $-0.0284\sym{***}$ & $0.0106\sym{***}$ \\
 & $(0.0030)$ & $(0.0025)$ & $(0.0039)$ & $(0.0006)$ \\
MiniLM XGBoost & $-0.0278\sym{***}$ & $-0.0198\sym{***}$ & $-0.0261\sym{***}$ & $0.0096\sym{***}$ \\
 & $(0.0030)$ & $(0.0023)$ & $(0.0036)$ & $(0.0007)$ \\
Arctic OLS & $-0.0273\sym{***}$ & $-0.0199\sym{***}$ & $-0.0226\sym{***}$ & $0.0099\sym{***}$ \\
 & $(0.0040)$ & $(0.0034)$ & $(0.0053)$ & $(0.0009)$ \\
\bottomrule
\end{tabular}
\vspace{0.35em}
\begin{flushleft}
\footnotesize Notes: Cells report $\text{Exposure}\times\text{Post}$ coefficients. Worker-level bootstrap standard errors are in parentheses. For the three secondary specifications, bootstrap repetitions use an $m$-out-of-$n$ worker bootstrap \citep{bickel1997resampling}, resampling 10\% of workers and scaling draw deviations by $\sqrt{m/N}$, where $m$ is the bootstrap worker sample size and $N$ is the full worker sample. \sym{*} $p<0.10$, \sym{**} $p<0.05$, \sym{***} $p<0.01$.
\end{flushleft}
\end{table}

\subsection{Pre-Trend Diagnostics}
\label{app:pretrends}

Table \ref{tab:pretrends} reports formal pre-trend diagnostics for the five main event-study outcomes. For each outcome, we test whether the six pre-treatment exposure interactions (2021Q1--2022Q2, relative to 2022Q3) are jointly zero, using a Wald statistic with $p$-values from the worker-level bootstrap. Following \citet{roth2022}, we report the root-mean-square of the pre-period coefficients alongside the tests, rather than relying only on statistical significance. This comparison helps assess the size of the pre-trend deviations relative to the coefficient of interest. 

\begin{table}[H]
\centering
\caption{Pre-Trend Diagnostics}
\label{tab:pretrends}
\vspace{0.4em}
\begin{tabular}{lccc}
\toprule
Outcome & Joint Wald $p$-Value & RMS Pre-Trend & Post-Period Estimate \\
\midrule
Self-Presentation & 0.28 & 0.0043 & $-0.0285$ \\
Credentials & 0.28 & 0.0036 & $-0.0238$ \\
Reputation & 0.61 & 0.0048 & $-0.0294$ \\
Price & 0.01 & 0.0017 & $0.0111$ \\
Observed Demand & 0.30 & 0.0176 & $-0.0725$ \\
\bottomrule
\end{tabular}
\vspace{0.35em}
\begin{flushleft}
\footnotesize Notes: The joint Wald test uses the six pre-treatment exposure interactions. Bootstrap $p$-values come from the worker-level bootstrap. RMS pre-trend is the root mean square of the six pre-period coefficients.
\end{flushleft}
\end{table}

\subsection{Mechanism Test with Price Controls}
\label{app:wage-control}

Because AI exposure varies across job categories that may also differ in price levels, one concern is that the demand-gap and demand-shift results in Figure \ref{fig:mechanism-ddd-panel} and Table \ref{tab:mechanism-pooled} partly reflect different post-AI demand paths for workers with different posted prices. Table \ref{tab:mechanism-wage-control} re-estimates the mechanism specification in Equation \eqref{eq:mechanism-ddd}, adding $P_i=\log(\text{posted hourly rate}_i)$ and period-specific price controls. In the post-period specification, the added terms are $P_i$ and $P_i\times \text{Post}_q$. In the late-post specification, the added terms are $P_i$, $P_i\times \text{EarlyPost}_q$, and $P_i\times \text{LatePost}_q$, where $\text{EarlyPost}_q$ covers 2022Q4--2025Q1 and $\text{LatePost}_q$ covers 2025Q2--2026Q1. These controls allow demand to shift differently after ChatGPT for workers with different posted prices, while the coefficients of interest continue to capture whether the high-minus-low demand gap changes more in more AI-exposed categories.

\begin{table}[H]
\centering
\caption{Demand Gap Compression with Price Controls}
\label{tab:mechanism-wage-control}
\vspace{0.4em}
\begin{tabular}{lcc}
\toprule
 & Human Capital & Price \\
\midrule
$\text{Top} \times \text{Exposure} \times \text{Post}$ & $-0.0512\sym{**}$ & $-0.0325$ \\
 & $(0.0209)$ & $(0.0225)$ \\
\addlinespace
$\text{Top} \times \text{Exposure} \times \text{LatePost}$ & $-0.0846\sym{***}$ & $-0.0818\sym{***}$ \\
 & $(0.0257)$ & $(0.0277)$ \\
\bottomrule
\end{tabular}
\vspace{0.35em}
\begin{flushleft}
\footnotesize Notes: Both rows re-estimate the mechanism specifications from Table \ref{tab:mechanism-pooled} with log posted hourly price and price-by-post controls. The late-post row includes unreported EarlyPost interactions for 2022Q4--2025Q1. Worker-clustered standard errors are in parentheses. \sym{*} $p<0.10$, \sym{**} $p<0.05$, \sym{***} $p<0.01$.
\end{flushleft}
\end{table}

The results are similar to the baseline mechanism estimates. In the post-period specification, the human capital estimate falls modestly from $-0.064$ to $-0.051$ and remains statistically significant, while the price estimate is essentially unchanged. In the late-post specification, both estimates remain large and statistically distinguishable from zero. These results suggest that the mechanism findings are not driven solely by different average post-period demand paths for workers at different posted prices.

\subsection{Cross-Sectional and Temporal Variability in Prices}
\label{app:price-variation}

The price block uses posted hourly rates from the March 2026 profile snapshot because historical posted profile rates are not observed. We do observe realized hourly rates for some contracts, although only for hourly contracts, which are 38.9\% of all contracts; the remaining 61.1\% are fixed-price contracts. In addition, the hourly rates we observe are realized prices, not posted prices, and posted prices are what clients initially see when screening workers. We therefore use the profile rate as the main price measure.

\begin{figure}[H]
\centering
\caption{Cross-Sectional and Temporal Variability in Prices}
\label{fig:snapshot-wage-group-trends}
\vspace{0.35em}
\includegraphics[width=0.90\textwidth]{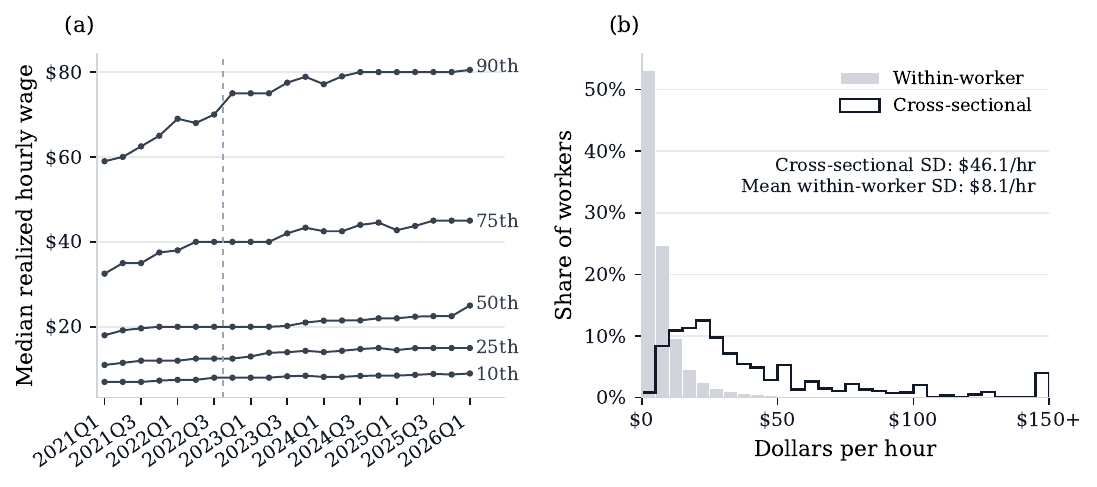}
\vspace{0.25em}
\begin{flushleft}
\footnotesize Notes: The left panel ranks workers once by March 2026 posted hourly rate and plots median realized hourly rates for bands centered on the 10th, 25th, 50th, 75th, and 90th percentiles. The dashed vertical line marks the first post-AI quarter. The right panel compares cross-sectional posted rates with within-worker standard deviations in realized hourly rates. The rightmost bin includes values at or above \$150 per hour.
\end{flushleft}
\end{figure}

Figure \ref{fig:snapshot-wage-group-trends} depicts realized hourly rates for workers at different percentiles of the price distribution. This diagnostic confirms that most of the variation in prices is cross-sectional rather than temporal: Workers ranked higher by the 2026 posted-rate snapshot also had higher realized hourly rates throughout the sample. Snapshot posted rates are strongly aligned with pre-AI realized hourly rates, with Pearson and rank correlations of 0.85. In addition, within-worker price variation is small relative to cross-sectional variation: The average within-worker standard deviation is \$8.1, compared with a snapshot posted-rate standard deviation of \$46.1 and an average quarterly cross-sectional realized-rate standard deviation of \$37.9. This suggests that the March 2026 snapshot provides a reasonable approximation of each worker's relative position in the platform's price distribution, and may also be close to the actual posted price clients observed at the time of hiring.

\subsection{Placebo Tests}
\label{app:permutation}

We randomly reassign AI exposure scores across the 102 job subcategories, holding each worker's subcategory fixed, and re-estimate the difference-in-differences specifications 2,000 times. This placebo exercise tests whether the observed estimates are unusual relative to random exposure assignments.

\begin{figure}[H]
\centering
\caption{Permutation Test for Block Importance}
\label{fig:permutation-placebo}
\vspace{0.35em}
\includegraphics[width=0.95\textwidth]{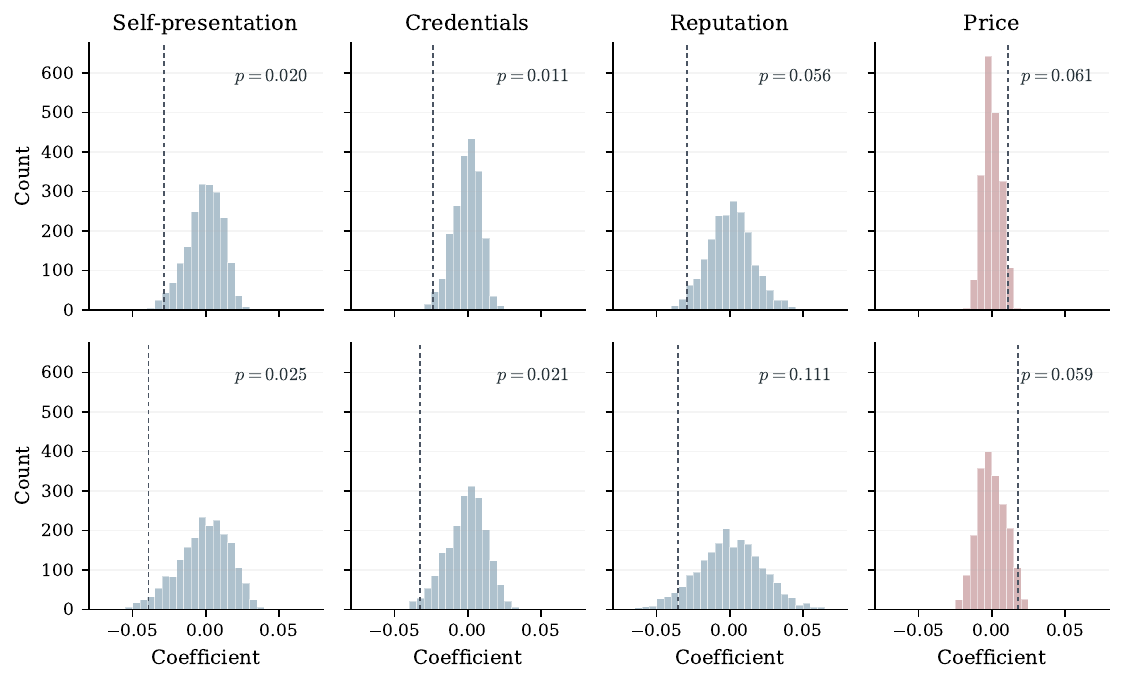}
\vspace{0.35em}
\begin{flushleft}
\footnotesize Notes: Histograms plot coefficients from the permutation tests. The top row reports the baseline post-period specification; the bottom row reports the late-post specification, showing the coefficient for the 2025Q2--2026Q1 period. Vertical lines mark the observed coefficients. The permutation $p$-value is the share of draws with an absolute coefficient at least as large as observed.
\end{flushleft}
\end{figure}

For the demand-component estimates, the observed post-period coefficients lie in the tails of the permutation distributions, with two-sided $p$-values between 0.011 and 0.061. The late-post coefficients are larger in magnitude, although the final-four-quarter permutation distributions are also wider; the corresponding two-sided $p$-values range from 0.022 to 0.111. The joint evidence is stronger: In both specifications, none of the 2,000 permutations reproduces declines in all three human capital components together with an increase in price at least as large as observed ($p<0.0005$).

\begin{figure}[H]
\centering
\caption{Permutation Test for Demand Gaps}
\label{fig:mechanism-permutation-placebo}
\vspace{0.35em}
\includegraphics[width=0.78\textwidth]{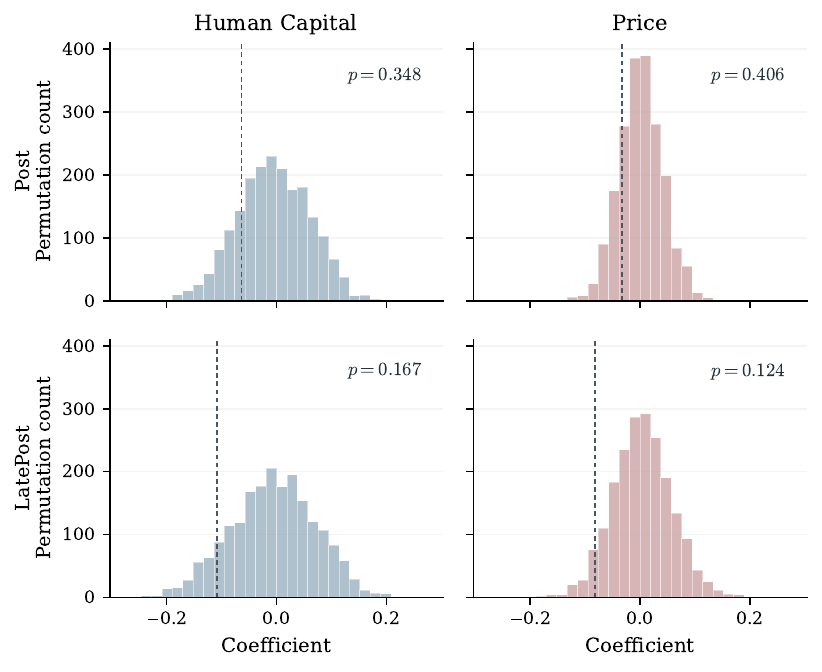}
\vspace{0.35em}
\begin{flushleft}
\footnotesize Notes: Histograms plot pooled triple-difference coefficients from the exposure-permutation exercise. The top row reports the baseline post-period specification; the bottom row reports the late-post specification for 2025Q2--2026Q1. Vertical lines mark the observed coefficients. Coefficients compare an unexposed category with the most exposed category. The plotted $p$-values are two-sided permutation $p$-values.
\end{flushleft}
\end{figure}

For the mechanism tests, the individual baseline coefficients are not extreme under the permutation distribution. The late-post coefficients are more unusual, with two-sided $p$-values of 0.167 for the human-capital demand-gap estimate and 0.124 for the price demand-shift estimate. The joint directional pattern provides stronger support: Only 4.4\% of baseline permutations produce both a human-capital demand-gap decline and a shift toward lower-priced workers at least as large as observed, and this share falls to 0.5\% for the late-post specification.

\clearpage

\section{Skill Tags}
\label{app:skill-tags}

Table \ref{tab:skill-tag-examples} reports common skill tags from worker profiles. These tags are part of the self-presentation block. In the main analysis, they are not hand-coded into separate variables; instead, they are embedded together with the worker's profile title and description.

\begingroup
\singlespacing
\setlength{\tabcolsep}{1.8pt}
\renewcommand{\arraystretch}{0.62}
\setlength{\extrarowheight}{0pt}
\setlength{\aboverulesep}{0pt}
\setlength{\belowrulesep}{0pt}
\setlength{\cmidrulesep}{0pt}
\setlength{\defaultaddspace}{0pt}
\setlength{\LTpre}{0pt}
\setlength{\LTpost}{0pt}
\begin{longtable}{@{}>{\fontsize{7}{7}\selectfont\raggedright\arraybackslash}p{0.22\textwidth}>{\fontsize{7}{7}\selectfont\raggedright\arraybackslash}p{0.74\textwidth}@{}}
\caption{Examples of Skill Tags}\label{tab:skill-tag-examples}\\[0.4em]
\toprule
Primary Category & Common Skill Tags \\
\midrule
\endfirsthead
\caption[]{Examples of Skill Tags (continued)}\\[0.4em]
\toprule
Primary Category & Common Skill Tags \\
\midrule
\endhead
\midrule
\multicolumn{2}{r}{\footnotesize Continued on next page} \\
\endfoot
\bottomrule
\endlastfoot
Accounting \& Consulting & Microsoft Excel; Data Entry; Bookkeeping; Intuit QuickBooks; Administrative Support; Bank Reconciliation; Customer Service; Xero; Project Management; Lead Generation \\
Admin Support & Data Entry; Lead Generation; Microsoft Excel; Administrative Support; Social Media Marketing; Social Media Management; Data Mining; Content Writing; Online Research; Graphic Design \\
Customer Service & Customer Service; Data Entry; Administrative Support; Customer Support; Email Support; Online Chat Support; Lead Generation; Email Communication; Virtual Assistance; Social Media Management \\
Data Science \& Analytics & Python; Data Analysis; Machine Learning; Data Visualization; SQL; Data Science; Microsoft Excel; Google Ads; Google Analytics; Microsoft Power BI \\
Design \& Creative & Adobe Photoshop; Graphic Design; Adobe Illustrator; Video Editing; Logo Design; Illustration; Adobe Premiere Pro; Adobe After Effects; Photo Editing; Adobe InDesign \\
Engineering \& Architecture & 3D Modeling; 3D Rendering; Autodesk AutoCAD; Adobe Photoshop; 3D Design; SketchUp; Interior Design; Architectural Design; CAD; 3D Animation \\
IT \& Networking & Data Entry; Lead Generation; Microsoft Excel; Data Mining; Data Scraping; List Building; Online Research; Administrative Support; Python; Email Marketing \\
Legal & Legal Research; Contract Drafting; Legal Writing; Legal Consulting; Contract Law; Corporate Law; Legal Assistance; Legal; Intellectual Property Law; Data Entry \\
Sales \& Marketing & Search Engine Optimization; Google Ads; SEO Keyword Research; Social Media Marketing; SEO Backlinking; SEO Audit; On-Page SEO; Copywriting; Off-Page SEO; Content Writing \\
Translation & Translation; Proofreading; English; Content Writing; Copywriting; General Transcription; Data Entry; Article Writing; Writing; Editing \& Proofreading \\
Web, Mobile \& Software Dev & WordPress; JavaScript; PHP; React; Web Design; Node.js; Web Development; Shopify; CSS; Python \\
Writing & Content Writing; Blog Writing; Copywriting; Article Writing; Writing; Creative Writing; Proofreading; English; Editing \& Proofreading; SEO Writing \\
\end{longtable}
\begin{flushleft}
\footnotesize Notes: The table reports the ten most common active skill tags among workers in the analysis sample, separately by primary Upwork category. Tags are counted once per worker within category. Skill tags enter the self-presentation block jointly with profile title and description.
\end{flushleft}
\endgroup

\section{AI-Exposure Scores by Job Subcategory}
\label{app:exposure-mapping}

\begin{figure}[H]
\centering
\caption{Distribution of AI Exposure Scores}
\label{fig:e1-exposure-distribution}
\vspace{0.35em}
\includegraphics[width=0.90\textwidth]{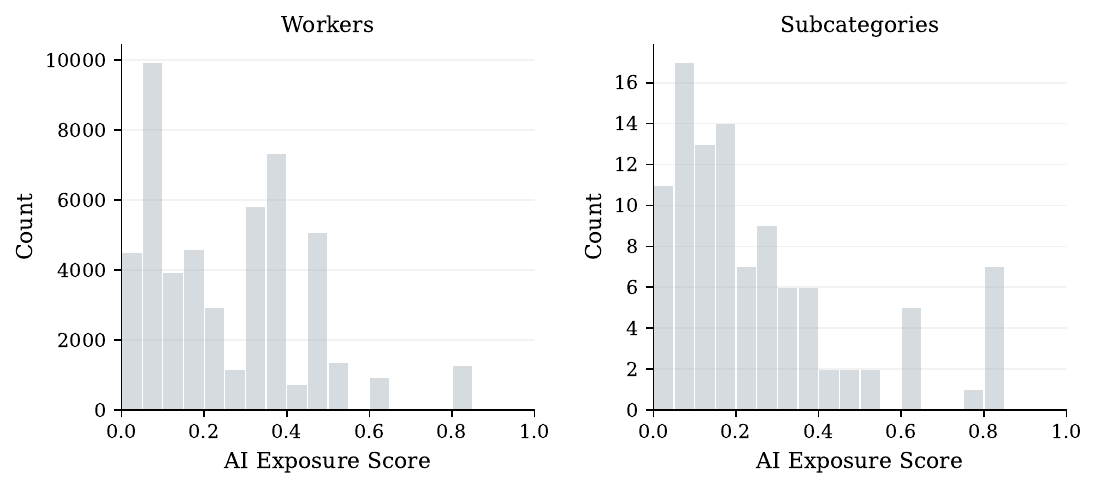}
\vspace{0.25em}
\begin{flushleft}
\footnotesize Notes: The worker distribution is based on each worker's primary selected Upwork subcategory; its mean is 0.252 and median is 0.219. The subcategory distribution weights each of the 102 represented Upwork subcategories equally; its mean is 0.252 and median is 0.186.
\end{flushleft}
\end{figure}

Figure \ref{fig:e1-exposure-distribution} reports the distribution of raw AI exposure scores, based on matching Upwork subcategories to occupational exposure scores from \citet{eloundou2024}. The left panel shows the distribution of worker-level exposure scores in our sample. The right panel reports the unweighted distribution across the 102 primary Upwork subcategories represented in the sample. The largest raw score across all subcategories is 0.8 (Translation). To simplify interpretation of our results, in our regressions we divide all raw exposure scores by 0.8, so coefficient can be interpreted as the difference between the most exposed category and the least. 

Table \ref{tab:upwork-onet-mapping} reports the occupation mapping used to assign AI exposure. The mapping links selected Upwork subcategories to O*NET occupations in the occupation-level file from \citet{eloundou2024}. We constructed the mapping using an LLM to match platform subcategory names to O*NET occupation titles, then assigned E1 scores from the matched occupations. 

\begingroup
\singlespacing
\setlength{\tabcolsep}{1.8pt}
\renewcommand{\arraystretch}{0.62}
\setlength{\extrarowheight}{0pt}
\setlength{\aboverulesep}{0pt}
\setlength{\belowrulesep}{0pt}
\setlength{\cmidrulesep}{0pt}
\setlength{\defaultaddspace}{0pt}
\setlength{\LTpre}{0pt}
\setlength{\LTpost}{0pt}
\setlength{\LTcapwidth}{\textwidth}
\begin{longtable}{@{}>{\fontsize{7}{7}\selectfont\raggedright\arraybackslash}p{0.38\textwidth}>{\fontsize{7}{7}\selectfont\raggedright\arraybackslash}p{0.48\textwidth}>{\fontsize{7}{7}\selectfont\raggedleft\arraybackslash}p{0.09\textwidth}@{}}
\caption{Upwork Subcategory to O*NET Occupation Mapping and AI Exposure}\label{tab:upwork-onet-mapping}\\[0.4em]
\toprule
Upwork Subcategory & O*NET Occupation & AI Exposure \\
\midrule
\endfirsthead
\caption[]{Upwork Subcategory to O*NET Occupation Mapping and AI Exposure (continued)}\\[0.4em]
\toprule
Upwork Subcategory & O*NET Occupation & AI Exposure \\
\midrule
\endhead
\midrule
\multicolumn{3}{r}{\footnotesize Continued on next page} \\
\endfoot
\bottomrule
\endlastfoot
3D Modeling \& CAD & Architectural and Civil Drafters & 0.180 \\
A/B Testing & Statisticians & 0.474 \\
AI \& Machine Learning & Data Scientists & 0.250 \\
AI Apps \& Integration & Software Developers & 0.053 \\
AI Integration & Software Developers & 0.053 \\
Academic Writing \& Research & Writers and Authors & 0.615 \\
Accounting & Accountants and Auditors & 0.200 \\
Accounting \& Bookkeeping & Bookkeeping, Accounting, and Auditing Clerks & 0.140 \\
Animation & Special Effects Artists and Animators & 0.050 \\
Architecture & Architects, Except Landscape and Naval & 0.163 \\
Art \& Illustration & Fine Artists, Including Painters, Sculptors, and Illustrators & 0.050 \\
Article \& Blog Writing & Writers and Authors & 0.615 \\
Audio \& Music Production & Sound Engineering Technicians & 0.000 \\
Audio Production & Sound Engineering Technicians & 0.000 \\
Automation Testing & Software Quality Assurance Analysts and Testers & 0.316 \\
Back-End Development & Software Developers & 0.053 \\
Blockchain, NFT \& Cryptocurrency & Blockchain Engineers & 0.235 \\
Brand Identity \& Strategy & Advertising and Promotions Managers & 0.256 \\
Branding & Advertising and Promotions Managers & 0.256 \\
Branding \& Logo Design & Graphic Designers & 0.000 \\
Building \& Landscape Architecture & Landscape Architects & 0.053 \\
Business \& Corporate Law & Lawyers & 0.150 \\
Chemical \& Process Engineering & Chemical Engineers & 0.077 \\
Chemical Engineering & Chemical Engineers & 0.077 \\
Civil \& Structural Engineering & Civil Engineers & 0.143 \\
Cloud Engineering & Computer Network Architects & 0.203 \\
Community Management \& Tagging & Customer Service Representatives & 0.545 \\
Consulting & Management Analysts & 0.190 \\
Content \& Copywriting & Writers and Authors & 0.615 \\
Content Writing & Writers and Authors & 0.615 \\
Contract Manufacturing & Industrial Production Managers & 0.152 \\
Copywriting & Writers and Authors & 0.615 \\
Corporate \& Contract Law & Lawyers & 0.150 \\
Corporate Law & Lawyers & 0.150 \\
Creative Writing & Poets, Lyricists and Creative Writers & 0.667 \\
Creative Writing Services & Poets, Lyricists and Creative Writers & 0.667 \\
Customer Experience \& Tech Support & Computer User Support Specialists & 0.481 \\
Customer Service & Customer Service Representatives & 0.545 \\
Customer Service \& Tech Support & Computer User Support Specialists & 0.481 \\
Data Analysis \& Testing & Data Scientists & 0.250 \\
Data Analytics & Data Scientists & 0.250 \\
Data Design \& Visualization & Data Scientists & 0.250 \\
Data Engineering & Database Architects & 0.244 \\
Data Entry & Data Entry Keyers & 0.214 \\
Data Entry \& Transcription Services & Data Entry Keyers & 0.214 \\
Data Extraction / ETL & Database Architects & 0.244 \\
Data Extraction/ETL & Database Architects & 0.244 \\
Data Mining \& Management & Data Scientists & 0.250 \\
Data Visualization & Data Scientists & 0.250 \\
Database Administration & Database Administrators & 0.290 \\
Database Development & Database Architects & 0.244 \\
Database Management \& Administration & Database Administrators & 0.290 \\
Desktop Application Development & Software Developers & 0.053 \\
Desktop Software Development & Software Developers & 0.053 \\
DevOps \& Solution Architecture & Computer Systems Engineers/Architects & 0.214 \\
DevOps \& Solutions Architecture & Computer Systems Engineers/Architects & 0.214 \\
DevOps Engineering & Computer Systems Engineers/Architects & 0.214 \\
Digital Marketing & Search Marketing Strategists & 0.016 \\
Display Advertising & Advertising Sales Agents & 0.436 \\
ERP / CRM Software & Computer Systems Analysts & 0.118 \\
ERP/CRM Software & Computer Systems Analysts & 0.118 \\
Ecommerce Development & Web Developers & 0.308 \\
Editing \& Proofreading & Editors & 0.529 \\
Editing \& Proofreading Services & Editors & 0.529 \\
Electrical \& Electronic Engineering & Electrical Engineers & 0.156 \\
Electrical \& Electronics Engineering & Electrical Engineers & 0.156 \\
Electrical Engineering & Electrical Engineers & 0.156 \\
Email \& Marketing Automation & Search Marketing Strategists & 0.016 \\
Energy \& Mechanical Engineering & Mechanical Engineers & 0.186 \\
Experimentation \& Testing & Statisticians & 0.474 \\
Finance & Financial and Investment Analysts & 0.154 \\
Finance \& Tax Law & Lawyers & 0.150 \\
Financial Planning & Personal Financial Advisors & 0.395 \\
Front-End Development & Web Developers & 0.308 \\
Full Stack Development & Software Developers & 0.053 \\
Game Design \& Development & Video Game Designers & 0.167 \\
Game Development & Video Game Designers & 0.167 \\
General Translation & Interpreters and Translators & 0.800 \\
Grant \& Proposal Writing & Writers and Authors & 0.615 \\
Graphic Design & Graphic Designers & 0.000 \\
Graphic, Editorial \& Presentation Design & Graphic Designers & 0.000 \\
Graphics \& Design & Graphic Designers & 0.000 \\
Hardware \& Firmware Development & Computer Hardware Engineers & 0.182 \\
Human Resources & Human Resources Specialists & 0.378 \\
Illustration & Fine Artists, Including Painters, Sculptors, and Illustrators & 0.050 \\
Immigration Law & Lawyers & 0.150 \\
Industrial Engineering & Industrial Engineers & 0.316 \\
Information Security & Information Security Analysts & 0.087 \\
Information Security \& Compliance & Information Security Analysts & 0.087 \\
Intellectual Property Law & Lawyers & 0.150 \\
Interior \& Trade Show Design & Interior Designers & 0.138 \\
Interior Design & Interior Designers & 0.138 \\
International \& Immigration Law & Lawyers & 0.150 \\
International Law & Lawyers & 0.150 \\
Language Localization & Interpreters and Translators & 0.800 \\
Language Tutoring \& Interpretation & Interpreters and Translators & 0.800 \\
Lead Generation & Telemarketers & 0.222 \\
Lead Generation \& Telemarketing & Telemarketers & 0.222 \\
Legal Translation & Interpreters and Translators & 0.800 \\
Legal, Medical \& Technical Translation & Interpreters and Translators & 0.800 \\
Logo Design \& Branding & Graphic Designers & 0.000 \\
Machine Learning & Data Scientists & 0.250 \\
Management Consulting & Management Analysts & 0.190 \\
Management Consulting \& Analysis & Management Analysts & 0.190 \\
Market \& Customer Research & Market Research Analysts and Marketing Specialists & 0.231 \\
Market Research \& Product Reviews & Market Research Analysts and Marketing Specialists & 0.231 \\
Marketing \& Brand Strategy & Marketing Managers & 0.219 \\
Marketing Strategy & Marketing Managers & 0.219 \\
Marketing, PR \& Brand Strategy & Marketing Managers & 0.219 \\
Mechanical Engineering & Mechanical Engineers & 0.186 \\
Medical Translation & Interpreters and Translators & 0.800 \\
Mobile Development & Software Developers & 0.053 \\
Motion Graphics & Special Effects Artists and Animators & 0.050 \\
NFT, AR/VR \& Game Art & Video Game Designers & 0.167 \\
Network \& System Administration & Network and Computer Systems Administrators & 0.162 \\
Online Research & Survey Researchers & 0.750 \\
Other - Accounting \& Consulting & Management Analysts & 0.190 \\
Other - Admin Support & Secretaries and Administrative Assistants, Except Legal, Medical, and Executive & 0.365 \\
Other - Customer Service & Customer Service Representatives & 0.545 \\
Other - Data Science \& Analytics & Data Scientists & 0.250 \\
Other - Design \& Creative & Graphic Designers & 0.000 \\
Other - Design \& Creative & Graphic Designers & 0.000 \\
Other - Engineering & Mechanical Engineers & 0.186 \\
Other - Engineering \& Architecture & Mechanical Engineers & 0.186 \\
Other - IT \& Networking & Computer Network Support Specialists & 0.231 \\
Other - Sales \& Marketing & Market Research Analysts and Marketing Specialists & 0.231 \\
Other - Software Development & Software Developers & 0.053 \\
Other - Writing & Writers and Authors & 0.615 \\
Other Customer Service \& Support & Customer Service Representatives & 0.545 \\
Paralegal & Paralegals and Legal Assistants & 0.050 \\
Paralegal Services & Paralegals and Legal Assistants & 0.050 \\
Performing Arts & Actors & 0.125 \\
Personal \& Professional Coaching & Training and Development Specialists & 0.421 \\
Personal / Virtual Assistant & Secretaries and Administrative Assistants, Except Legal, Medical, and Executive & 0.365 \\
Personal/Virtual Assistance & Secretaries and Administrative Assistants, Except Legal, Medical, and Executive & 0.365 \\
Photography & Photographers & 0.021 \\
Physical Design & Commercial and Industrial Designers & 0.069 \\
Physical Sciences & Chemists & 0.280 \\
Presentations & Graphic Designers & 0.000 \\
Product \& Physical Design & Commercial and Industrial Designers & 0.069 \\
Product Design & Commercial and Industrial Designers & 0.069 \\
Product Design & Commercial and Industrial Designers & 0.069 \\
Product Management & Project Management Specialists & 0.050 \\
Product Management \& Scrum & Project Management Specialists & 0.050 \\
Professional \& Business Writing & Writers and Authors & 0.615 \\
Project \& Product Management & Project Management Specialists & 0.050 \\
Project Management & Project Management Specialists & 0.050 \\
Public Law & Lawyers & 0.150 \\
Public Relations & Public Relations Specialists & 0.636 \\
QA \& Testing & Software Quality Assurance Analysts and Testers & 0.316 \\
QA Testing & Software Quality Assurance Analysts and Testers & 0.316 \\
Quantitative Analysis & Financial Quantitative Analysts & 0.324 \\
Recruiting \& Human Resources & Human Resources Specialists & 0.378 \\
Regulatory Law & Lawyers & 0.150 \\
Renewable Energy Engineering & Environmental Engineers & 0.241 \\
Resumes \& Cover Letters & Writers and Authors & 0.615 \\
SEM - Search Engine Marketing & Search Marketing Strategists & 0.016 \\
SEO \& SEM Services & Search Marketing Strategists & 0.016 \\
SEO - Search Engine Optimization & Search Marketing Strategists & 0.016 \\
SMM - Social Media Marketing & Search Marketing Strategists & 0.016 \\
Sales \& Business Development & Sales Representatives of Services, Except Advertising, Insurance, Financial Services, and Travel & 0.333 \\
Sales \& Marketing Copywriting & Writers and Authors & 0.615 \\
Scripting \& Automation & Software Developers & 0.053 \\
Scripts \& Utilities & Computer Programmers & 0.433 \\
Securities \& Finance Law & Lawyers & 0.150 \\
Security \& Compliance & Information Security Analysts & 0.087 \\
Social Media \& PR Services & Public Relations Specialists & 0.636 \\
Social Media Marketing \& Strategy & Search Marketing Strategists & 0.016 \\
Systems Engineering \& Architecture & Computer Systems Engineers/Architects & 0.214 \\
Tax Law & Lawyers & 0.150 \\
Tech Support \& Content Moderation & Computer User Support Specialists & 0.481 \\
Technical Support & Computer User Support Specialists & 0.481 \\
Technical Translation & Interpreters and Translators & 0.800 \\
Technical Writing & Technical Writers & 0.333 \\
Telemarketing \& Telesales & Telemarketers & 0.222 \\
Transcription & Medical Transcriptionists & 0.143 \\
Translation & Interpreters and Translators & 0.800 \\
Translation \& Localization & Interpreters and Translators & 0.800 \\
Translation \& Localization Services & Interpreters and Translators & 0.800 \\
Video \& Animation & Film and Video Editors & 0.139 \\
Video Production & Producers and Directors & 0.255 \\
Virtual Assistance & Secretaries and Administrative Assistants, Except Legal, Medical, and Executive & 0.365 \\
Virtual/Administrative Assistance & Secretaries and Administrative Assistants, Except Legal, Medical, and Executive & 0.365 \\
Voice Talent & Actors & 0.125 \\
Web \& Mobile Design & Web and Digital Interface Designers & 0.367 \\
Web Content & Writers and Authors & 0.615 \\
Web Development & Web Developers & 0.308 \\
Web Research & Survey Researchers & 0.750 \\
Wordpress \& CMS & Web Developers & 0.308 \\
\end{longtable}
\begin{flushleft}
\footnotesize Notes: The table reports the mapping used to assign AI exposure to freelancers. Each freelancer is assigned exposure using the first matched selected Upwork subcategory on the profile. The final column reports the raw E1 exposure score from \citet{eloundou2024}.
\end{flushleft}
\endgroup

\end{document}